\documentclass[a4paper,11pt]{article}
\pdfoutput=1 
\usepackage{jcappub}
\usepackage{graphicx}
\usepackage{amsmath}
\usepackage{amssymb}
\usepackage{amsfonts}
\usepackage{mathrsfs}
\usepackage{lscape}
\usepackage{verbatim}
\usepackage{xcolor}
\usepackage{bm}
\usepackage{tabularx}
\usepackage{mathtools}
\usepackage{ulem}
\usepackage{hyperref}
\newcommand{\ve}{\varepsilon}
\newcommand{\mc}{\mathcal}

\renewcommand{\(}{\left(}
\renewcommand{\)}{\right)}
\renewcommand{\[}{\left[}
\renewcommand{\]}{\right]}
\definecolor{darkgreen}{rgb}{0,0.6,0}
\makeatother

\title{Tracking the validity of the quasi-static and sub-horizon approximations in modified gravity}

\author[\star, 1]{J. Bayron Orjuela-Quintana,\note{Corresponding author.}}
\author[\ddagger, 2]{Savvas Nesseris,}

\affiliation[1]{Departamento de F\'isica, Universidad del Valle, Ciudad Universitaria Mel\'endez, 760032, Cali, Colombia}

\affiliation[2]{Instituto  de  F\'isica  Te\'orica  UAM-CSIC,  Universidad  Auton\'oma  de  Madrid,  Cantoblanco,  28049  Madrid,  Spain}

\emailAdd{$\star$ john.orjuela@correounivalle.edu.co}
\emailAdd{\mbox{$\dagger$ savvas.nesseris@csic.es}}
\date{\today}

\abstract{Within the framework of modified gravity, the quasi-static and sub-horizon approximations are widely used in analyses aiming to identify departures from the concordance model at late-times. In general, it is assumed that time derivatives are subdominant with respect to spatial derivatives given that the relevant physical modes are those well inside the Hubble radius. In practice, the perturbation equations under these approximations are reduced to a tractable algebraic system in terms of the gravitational potentials and the perturbations of involved matter fields. Here, in the framework of $f(R)$ theories, we revisit standard results when these approximations are invoked using a new parameterization scheme that allows us to track the relevance of each time-derivative term in the perturbation equations. This new approach unveils correction terms which are neglected in the standard procedure. We assess the relevance of these differences by comparing results from both approaches against full numerical solutions for two well-known toy-models: the designer $f(R)$ model and the Hu-Sawicki model. We find that: $i)$ the sub-horizon approximation can be safely applied to linear perturbation equations for scales $0.06 \, h/\text{Mpc} \lesssim k \lesssim 0.2 \, h/\text{Mpc}$, $ii)$ in this ``safety region'', the quasi-static approximation provides a very accurate description of the late-time cosmological dynamics even when dark energy significantly contribute to the cosmic budget, and $iii)$ our new methodology performs better than the standard procedure, even for several orders of magnitude in some cases. Although, the impact of this major improvement on the linear observables is minimal for the studied cases, this does not represent an invalidation for our approach. Instead, our findings indicate that the perturbation expressions derived under these approximations in more general modified gravity theories, such as Horndeski, should be also revisited.}

\begin{document}

\maketitle

%%%%%%%%%%%%%%%%%%%%%%%%%%%%
\section{Introduction}
\label{Sec: Introduction}
%%%%%%%%%%%%%%%%%%%%%%%%%%%%

The current theoretical paradigm, namely the cosmological constant $\Lambda$ and cold dark matter (CDM) model $-$ $\Lambda$CDM \cite{Peebles2020}, has started to be challenged by several observations \cite{DiValentino:2020zio, DiValentino:2020vvd, Abdalla:2022yfr}. The well-known $H_0$ tension is now approximately $5\sigma$ \cite{Planck:2018vyg, Riess:2019cxk, Riess:2021jrx, Wong:2019kwg, Freedman:2021ahq}, while the $S_8$ tension is about $2-3\sigma$ \cite{BOSS:2016wmc, Heymans:2020gsg, KiDS:2021opn, DES:2022xxr, Macaulay:2013swa, Battye:2014qga, Ata:2017dya, Philcox:2021kcw, Kobayashi:2021oud, Huang:2021tvo, Blanchard:2021dwr}. Recently, the inconsistency between the amplitude of the Cosmic Microwave Background (CMB) dipole and the Quasars dipole has exceeded $5\sigma$, setting serious concerns on the cosmological principle \cite{Secrest:2020has, Dalang:2021ruy, Secrest:2022uvx, Dam:2022wwh}. These discrepancies and others (see Refs. \cite{Fields:2011zzb, Cyburt:2015mya, Pitrou:2020etk}) motivate a further examination of the viability of $\Lambda$CDM as the standard cosmological model \cite{Perivolaropoulos:2021jda, Euclid:2021frk, Euclid:2022ucc} and the exploration of models violating the cosmological principle \cite{Perivolaropoulos:2014lua, Garcia-Bellido:2008vdn, Garcia-Bellido:2008sdt, Redlich:2014gga, Orjuela-Quintana:2020klr, Guarnizo:2020pkj, Motoa-Manzano:2020mwe, Orjuela-Quintana:2021zoe, BeltranAlmeida:2021ywl}.

Generally, the main alternatives to the $\Lambda$CDM model imply either the existence of a fluid with negative pressure known as Dark Energy (DE) \cite{Copeland:2006wr, Amendola:2015ksp}, or covariant modifications to the geometric sector describing gravity, the so-called Modified Gravity (MG) theories \cite{Clifton:2011jh, CANTATA:2021ktz}. With respect to MG theories, a large portion in the plethora of candidate models are rule out by observations from, for example, gravitational waves \cite{LIGOScientific:2017ycc, Ezquiaga:2017ekz}, and very precise extra-galactic tests \cite{Planck:2015bue, LIGOScientific:2016lio, Collett:2018gpf}. However, there remains some alternatives. Among the still cosmologically viable MG theories, one of the most representative are the so-called $f(R)$ theories \cite{DeFelice:2010aj}. In principle, these theories could be distinguished from the standard cosmological model by their influence in the inhomogeneous Universe since, in general, they introduce modifications in several observables, e.g., the growth rate \cite{Pogosian:2007sw}. However, no $f(R)$ model is particularly favored by current growth data \cite{Perez-Romero:2017njc, AlvarezLuna:2018gzv}.

It is fair to say that the distinction between DE and MG is rather academic in a sense, as both scenarios can be treated on equal footing under the Effective Fluid Approach (EFA) \cite{Nesseris:2022hhc}.\footnote{The Effective Field Theory is another approach which allows to model dark energy and modified gravity in a common framework \cite{Gubitosi:2012hu, Bloomfield:2012ff, Hu:2013twa, Frusciante:2019xia}.} Within the EFA, all the modifications to gravity are encompassed in terms of the equation of state $w_\text{DE}$, the sound speed $c_{s, \text{DE}}^2$, and the anisotropic stress $\pi_\text{DE}$ of an effective DE fluid. As shown in Refs. \cite{Arjona:2018jhh, Arjona:2019rfn, Cardona:2022lcz}, the application of the Quasi-Static Approximation (QSA) and the Sub-Horizon Approximation (SHA) allows to get analytical expressions for the effective DE perturbations in very general MG theories. One of the advantage of this approach is that these expressions can be easily introduced in Boltzmann solvers, such as \texttt{CLASS} \cite{Blas:2011rf}, facilitating the computation of cosmological observables, e.g., the linear matter power spectrum or the CMB angular power spectrum. These approximations also enable us to directly discriminate departures from General Relativity (GR), which could be encoded in, for instance, the effective Newton's constant $G_\text{eff}$ and the effective lensing parameter $Q_\text{eff}$. Albeit they have been extensively used in the literature \cite{Tsujikawa:2007gd, Noller:2013wca, Silvestri:2013ne, Bose:2014zba, Bellini:2014fua, Gleyzes:2014rba, Sawicki:2015zya, Pogosian:2016pwr, DeFelice:2016uil, Pace:2020qpj}, some works claim that the QSA, in particular, can be too aggressive and that relevant dynamical information could be omitted after their application \cite{delaCruz-Dombriz:2008ium, Llinares:2013jua}. Therefore, the validity regime of these approximations is a highly relevant issue to be addressed in order to compare theoretical results with upcoming astrophysical observations.

Bearing in mind the above discussion, we propose to analyze the validity regime in time and scale of the QSA-SHA for modified gravity theories. We start our research program by revisiting well-known expressions obtained for the effective DE perturbations in $f(R)$ theories. In order to do so, we put forward a new parameterization scheme that allows us to track the relevance of each derivative term in the perturbation equations. In general, we find some correction terms which improve over the standard description of the effective DE perturbations by several orders of magnitude in some cases. 

This paper is organized as follows: in Sec.~\ref{Sec: Theoretical Framework}, we present the theoretical framework of linear cosmological perturbations for the late-time Universe around a Friedmann-Lema\^itre-Robertson-Walker (FLRW) metric. In Sec.~\ref{Sec: f(R) Theories}, the EFA is applied to $f(R)$ theories, and the standard QSA-SHA procedure is used to obtain well-known expressions for the gravitational potentials and the effective DE perturbations. Then, in Sec.~\ref{Sec: Tracking the Accuracy of the QSA and the SHA}, we introduce the new parameterization of the QSA-SHA and write down the dynamics in terms of these parameters. Since these parameters depend on the cosmological model, we introduce some specific $f(R)$ models in Sec.~\ref{Sec: Specific Models}. Then, in Sec.~\ref{Sec: Numerical Solution}, we compare the results obtained via the new parameterization, the standard QSA-SHA procedure, and the full numerical solution when no approximations are used. Since some differences between the standard approach and the new parameterization are found, we investigate the impact of these differences on some cosmological observables in \ref{Sec: Impact on Cosmological Observables}. We give a summary and our conclusions in Sec.~\ref{Sec: Conclusions}.

%%%%%%%%%%%%%%%%%%%%%%%%%%%%%%%%%%%%%%%%%%%%%%%
\section{Theoretical Framework}
\label{Sec: Theoretical Framework}
%%%%%%%%%%%%%%%%%%%%%%%%%%%%%%%%%%%%%%%%%%%%%%%

%%%%%%%%%%%%%%%%%%%%%%%%%%%%%%%%
\subsection{General Setup}
\label{SubSec: General Setup}
%%%%%%%%%%%%%%%%%%%%%%%%%%%%%%%%

In GR, gravitational interactions are characterized by the Einstein-Hilbert action
\begin{equation}
\label{Eq: EH action}
S_\text{EH} = \int \text{d}^4 x \sqrt{-g} \[ \frac{1}{2\kappa} R + \mc{L} \],
\end{equation}
where\footnote{We have set the speed of light to $c = 1$.} $\kappa \equiv 8 \pi G_\mathrm{N}$, $G_\mathrm{N}$ is the Newton's constant, $g$ is the determinant of the metric $g_{\mu\nu}$, $R$ is the Ricci scalar, and $\mc{L}$ is a Lagrangian for all matter fields. On cosmological scales, the Universe is approximately homogeneous and isotropic \cite{Planck:2018vyg}. Hence, its geometry can be described by the flat perturbed FLRW metric that in the Newtonian gauge reads
\begin{equation}
\label{Eq: FLRW metric}
\text{d} s^2 = - \{ 1 + 2 \Psi\( \boldsymbol{x}, t \) \} \text{d} t^2 + a(t)^2 \{ 1 + 2 \Phi \(\boldsymbol{x}, t \) \} \delta_{i j} \text{d} x^i \text{d} x^j,
\end{equation}
where $a$ is the scale factor, and $\Psi$ and $\Phi$ are the gravitational potentials which depend on both the spatial coordinates $\boldsymbol{x}$ and the cosmic time $t$. The dynamics is encoded in the Einstein field equations
\begin{equation}
\label{Eq: Einstein Eqs}
G^\mu_{\ \nu} = \kappa\,T^\mu_{\ \nu}, \qquad T^\mu_{\ \nu} = \( \rho + P \) u^\mu u_\nu + P \delta^\mu_{\ \nu},
\end{equation}
where $G^\mu_{\ \nu}$ is the Einstein tensor, $T^\mu_{\ \nu}$ is the energy-momentum tensor which is given in terms of the density $\rho$, the pressure $P$, and the velocity four-vector $u^\mu = (1 - \Psi, \frac{v^i}{a})$, with\footnote{An over-dot denotes a derivative with respect to $t$.} $v^i = a(t) \dot{x}^i$. The components of the energy-momentum tensor are given by
\begin{equation}
T^0_{\ 0} = - \(\bar{\rho} + \delta \rho \), \quad T^0_{\ i} = a(t) \( \bar{\rho} + \bar{P} \) i k_i v, \quad 
T^i_{\ j} = \( \bar{P} + \delta P \) \delta^i_{\ j} + \Sigma^i_{\ j},
\end{equation}
where $\bar{\rho}$ and $\bar{P}$ are the background density and pressure, $\delta \rho$ and $\delta P$ are their respective perturbations, $v$ defined from $v_i \equiv i k_i v$ is the velocity perturbation, $k_i$ is the wave-vector in Fourier space, and $\Sigma^i_j$ is the anisotropic stress tensor.

%%%%%%%%%%%%%%%%%%%%%%%%%%%%%%%%%%%%%%%%%%%%%%%
\subsection{Background and Perturbations} 
\label{SubSec: Background and Perturbations} 
%%%%%%%%%%%%%%%%%%%%%%%%%%%%%%%%%%%%%%%%%%%%%%%

Since we are interested in the late-time cosmological dynamics, we assume that the cosmic fluid is solely composed by pressure-less matter and a dark energy component with equation of state\footnote{The subscripts $m$ and DE means that the respective quantity is associated to matter or dark energy.} $\bar{P}_\text{DE} \equiv w_\text{DE} \bar{\rho}_\text{DE}$, where $w_\text{DE}$ depends on time in general.  This model is known as $w$CDM, and $\Lambda$CDM can be viewed as the particular case when $w_\text{DE} = -1$. For the background metric  in Eq.~\eqref{Eq: FLRW metric}, the unperturbed Einstein field equations \eqref{Eq: Einstein Eqs} give rise to the Friedmann equations
\begin{equation}
\label{Eq: Friedman Eqs}
H^2 = \frac{\kappa}{3} \( \bar{\rho}_m + \bar{\rho}_\text{DE} \), \quad \dot{H} = - \frac{\kappa}{2} \[ \bar{\rho}_m + \bar{\rho}_\text{DE} (1 + w_\text{DE}) \],
\end{equation}
where $H \equiv \dot{a} / a$ is the Hubble parameter. The linearized Einstein equations read
\begin{equation}
\label{Eq: Einstein 00}
\frac{k^2}{a^2} \Phi + 3 H \( \dot{\Phi} - H \Psi \) = \frac{\kappa}{2} \( \bar{\rho}_m \delta_m + \bar{\rho}_\text{DE} \delta_\text{DE} \),
\end{equation}
\begin{equation}
k^2 \( \dot{\Phi} - H \Psi \) = - \frac{\kappa}{2} a \[ \bar{\rho}_m \theta_m + \bar{\rho}_\text{DE} (1 + w_\text{DE}) \theta_\text{DE} \],
\end{equation}
\begin{equation}
\ddot{\Phi} + H \( 3 \dot{\Phi} - \dot{\Psi} \) - \( 2 \dot{H} + 3 H^2 \) \Psi + \frac{k^2}{3 a^2} \( \Phi + \Psi \) 
= - \frac{\kappa}{2} \delta P_\text{DE},
\end{equation}
\begin{equation}
\label{Eq: Traceless Einstein ij}
- \frac{k^2}{a^2} \( \Phi + \Psi \) = \frac{3 \kappa}{2} \bar{\rho}_\text{DE} (1 + w_\text{DE}) \sigma_\text{DE},
\end{equation}
corresponding to the ``time-time'', longitudinal ``time-space'', trace ``space-space'', and longitudinal
trace-less ``space-space'' parts of the gravitational field equations \eqref{Eq: Einstein Eqs}, respectively. In the later equations we have defined the following quantities: the perturbation over-density or density contrast $\delta \equiv \delta \rho / \bar{\rho}$, the velocity divergence $\theta \equiv i k^i v_i = - k^2 v$, and the scalar anisotropic stress $(\bar{\rho} + \bar{P}) \sigma \equiv - (\hat{k}_i \hat{k}_j - \frac{1}{3} \delta_{i j} ) \Sigma^{i j}$, where $k^2 \equiv k_i k^i$ and $\hat{k}_i \equiv k_i / k$. We have assumed that matter is also pressure-less at first-order ($\delta P_m = 0$) and that it has no anisotropic stress ($\sigma_m = 0$).

The conservation of the energy-momentum tensor, namely $\nabla_\mu T^\mu_{\ \nu} = 0$, impose the following zero-order continuity equations for matter and DE 
\begin{equation}
\label{Eq: Continuity Eqs}
\dot{\bar{\rho}}_m + 3 H \bar{\rho}_m = 0, \quad \dot{\bar{\rho}}_\text{DE} + 3 H (1 + w_\text{DE}) \bar{\rho}_\text{DE} = 0.
\end{equation}
At linear order we have:
\begin{equation}
\label{Eq: dm Eq}
\delta_m' = - 3 \Phi' - \frac{V_m}{a^2 H},
\end{equation}
\begin{equation}
\label{Eq: Vm Eq}
V_m' = - \frac{V_m}{a} + \frac{k^2}{a^2 H} \Psi,
\end{equation}
\begin{equation}
\delta_\text{DE}' = - 3 \( 1 + w_\text{DE} \) \Phi' - \frac{V_\text{DE}}{a^2 H} - \frac{3}{a} \( c_{s, \text{DE}}^2 - w_\text{DE} \) \delta_\text{DE},
\end{equation}
\begin{equation}
\label{Eq: DE V Eq}
V_\text{DE}' = - \(1 - 3w_\text{DE} \) \frac{V_\text{DE}}{a} + \frac{k^2}{a^2 H} c_{s, \text{DE}}^2 \delta_\text{DE} + \(1 + w_\text{DE} \) \frac{k^2}{a^2 H} \Psi - \frac{2}{3} \frac{k^2}{a^2 H} \pi_\text{DE},
\end{equation}
where we have defined the scalar velocity $V \equiv (1 + w) \theta$, the sound speed $c_s^2 \equiv \frac{\delta P}{\delta \rho}$, and the anisotropic stress parameter $\pi \equiv \frac{3}{2}(1 + w) \sigma$.\footnote{A prime denotes a derivative with respect to the scale factor.} 

In brief, to know the evolution of the large scale structure inhomogeneities of the Universe requires to solve the coupled set of differential equations given by the Einstein equations and the continuity equations, both at background and at linear order. As we will see in the next section, this task can be highly simplified for $f(R)$ theories under some well-motivated assumptions, namely the quasi-static and the sub-horizon approximations.

%%%%%%%%%%%%%%%%%%%%%%%%%%%%
\section{$f(R)$ Theories}
\label{Sec: f(R) Theories}
%%%%%%%%%%%%%%%%%%%%%%%%%%%%

In $f(R)$ models, covariant modifications of GR are performed by replacing $R$ in the Einstein-Hilbert action in Eq.~\eqref{Eq: EH action} by a general function $f(R)$ such that
\begin{equation}
\label{Eq: f(R) action}
S = \int \text{d}^4 x \sqrt{-g} \[ \frac{1}{2\kappa} f(R) + \mc{L} \].
\end{equation}
Upon varying this action with respect to the metric $g^{\mu\nu}$, the Einstein equations in Eq.~\eqref{Eq: Einstein Eqs} are replaced by
\begin{equation}
\label{Eq: f(R) field Eqs}
F G_{\mu\nu} - \frac{1}{2} \( f - F R \) g_{\mu\nu} + \(g_{\mu\nu} \Box - \nabla_\mu \nabla_\nu \) F = \kappa T_{\mu\nu},
\end{equation}
where\footnote{Derivatives with respect to $R$ are denoted using a sub-index, e.g., $\frac{\text{d}f}{\text{d}R} \equiv f_R$.} $F \equiv f_{R}$. Yet, it is possible to recast Eq.~\eqref{Eq: f(R) field Eqs} to look as the usual Einstein field equations \eqref{Eq: Einstein Eqs} using the effective fluid approach. Under this framework, all the modifications to gravity are encoded in a generalized DE fluid which, in the case of $f(R)$ theories, is characterized by the following energy-momentum tensor
\begin{equation}
\kappa T_{\mu\nu}^{\( \text{DE} \)} = \(1 - F \) G_{\mu\nu} + \frac{1}{2} \( f - F R \) g_{\mu\nu} - \( g_{\mu\nu} \Box - \nabla_\mu \nabla_\nu \) F. \label{Eq: f(R) energy tensor}
\end{equation}
Hence, the Friedmann equations are given by Eqs.~\eqref{Eq: Friedman Eqs} replacing the DE density and pressure by
\begin{align}
\kappa \bar{\rho}_\text{DE} &= - \frac{f}{2} + 3 H^2 \( 1 + F \) + 3 F \dot{H} - 3 H \dot{F}, \label{Eq: f(R) density} \\
\kappa \bar{P}_\text{DE} &= \frac{f}{2} - 3 H^2 \( 1 + F \) - \dot{H} \( 2 + F \) + 2 H \dot{F} + \ddot{F}. \label{Eq: f(R) pressure}
\end{align}
On the other hand, linear perturbation equations in Eqs.~\eqref{Eq: Einstein 00}-\eqref{Eq: Traceless Einstein ij} are modified as 
\begin{align}
0 &= - \kappa \bar{\rho}_m \delta_m + A_1 \dot{\Phi} + A_2 \dot{\Psi} + \( A_3 + A_4 \frac{k^2}{a^2} \) \Phi + \( A_5 + A_6 \frac{k^2}{a^2} \) \Psi, \label{Eq: 00 f(R)}\\
0 &= - \frac{a \kappa \bar{\rho}_m V_m}{k^2} + C_1 \dot{\Phi} + C_2 \dot{\Psi} + C_3 \Phi + C_4 \Psi,  \label{Eq: 0i f(R)} \\
0 &= B_1 \ddot{\Phi} + B_2 \ddot{\Psi} + B_3 \dot{\Phi} + B_4 \dot{\Psi} + B_5 \Phi + B_6 \Psi, \\
0 &= D_1 \ddot{\Phi} + D_2 \dot{\Phi} + D_3 \dot{\Psi} + \( D_4 + D_5 \frac{k^2}{a^2} \) \Phi + \( D_6 + D_7 \frac{k^2}{a^2} \) \Psi. \label{Eq: Traceless f(R)}
\end{align}
The last equation is obtained from the relation
\begin{equation}
\label{Eq: Usual traceless Eq}
0 = F_R (\Phi + \Psi) + F \delta R,
\end{equation}
where $\delta R$ is the perturbation of the Ricci scalar which in the metric \eqref{Eq: FLRW metric} reads
\begin{equation}
\label{Eq: dR}
\delta R = 6 \ddot{\Phi} + 24 H \dot{\Phi} - 6 H \dot{\Psi} + 4\frac{k^2}{a^2} \Phi + \( 2\frac{k^2}{a^2} - 12 \dot{H} + 24H^2 \) \Psi.
\end{equation}
The coefficients $A_i$, $C_i$, $B_i$, and $D_i$ can be found in Appendix \ref{App: Linear Field Equations in $f(R)$}. In the EFA, we can assign to the $f(R)$ theory an effective DE perturbation variable for density, pressure, velocity, and anisotropic stress respectively. They are obtained from the energy tensor in Eq.~\eqref{Eq: f(R) energy tensor} as
\begin{align}
\kappa \delta \rho_\text{DE} &= W_1 \dot{\Phi} + W_2 \dot{\Psi} + \( W_3 + W_4 \frac{k^2}{a^2} \) \Phi + \( W_5 + W_6 \frac{k^2}{a^2}\) \Psi, \label{Eq: deltaDE f(R)}\\
\kappa \delta P_\text{DE} &= Y_1 \ddot{\Phi} + Y_2 \ddot{\Psi} + Y_3 \dot{\Phi} + Y_4 \dot{\Psi} + \( Y_5 + Y_6 \frac{k^2}{a^2} \) \Phi + \( Y_7 + Y_8 \frac{k^2}{a^2} \) \Psi, \\
\frac{a \kappa \bar{\rho}_\text{DE}}{k^2} V_\text{DE} &= Z_1 \dot{\Phi} + Z_2 \dot{\Psi} + Z_3 \Phi + Z_4 \Psi, \label{Eq: VDE f(R)} \\
\kappa \bar{\rho}_\text{DE} \pi_\text{DE} &= - \frac{k^2}{a^2} \( \Phi + \Psi \), \label{Eq: stress f(R)}
\end{align}  
where the coefficients $W_i$, $Y_i$, and $Z_i$ can be found in Appendix \ref{App: Dark Energy Perturbations in $f(R)$}.

%%%%%%%%%%%%%%%%%%%%%%%%%%%%%%%%%%%%%%%%%%%%%%%%%%%%%%%%%%%%%%%%%%
\subsection{Quasi-Static and Sub-Horizon Approximations: Standard Approach}
%%%%%%%%%%%%%%%%%%%%%%%%%%%%%%%%%%%%%%%%%%%%%%%%%%%%%%%%%%%%%%%%%%

The dynamics of DE and matter inhomogeneities in $f(R)$ models can be unveiled if the evolution of the gravitational potentials is known. However, this implies to solve Eqs.~\eqref{Eq: 00 f(R)}-\eqref{Eq: Traceless f(R)} coupled to the matter perturbations equations in Eqs.~\eqref{Eq: dm Eq} and \eqref{Eq: Vm Eq}, which in general is a difficult task. As explained in Sec.~\ref{Sec: Introduction}, the application of the Quasi-Static Approximation (QSA) and the Sub-Horizon Approximation (SHA) can greatly simplify the related equations. In the following, we describe the standard steps to apply these approximations to the perturbation equations. We disregard time-derivatives of the potentials and consider that relevant modes for the cosmological dynamics are those well-inside the Hubble radius, i.e., $k \gg aH$. This allows us to neglect terms of the form $H \times$perturbation in favor of terms of the form $k^2 \times$perturbation. As an example, applying these approximations to the perturbation of the Ricci scalar in Eq.~\eqref{Eq: dR} gives
\begin{align*}
\delta R &= 6 \ddot{\Phi} + 24 H \dot{\Phi} - 6 H \dot{\Psi} + 4\frac{k^2}{a^2} \Phi + \( 2\frac{k^2}{a^2} - 12 \dot{H} + 24H^2 \) \Psi \\
 &\overset{\mathrm{QSA}}{\approx} 4\frac{k^2}{a^2} \Phi + \( 2\frac{k^2}{a^2} - 12 \dot{H} + 24H^2 \)\Psi \\
 &\overset{\mathrm{SHA}}{\approx} 4\frac{k^2}{a^2} \Phi + 2\frac{k^2}{a^2} \Psi.
\end{align*}
Using the QSA-SHA in the linear equations \eqref{Eq: 00 f(R)} and \eqref{Eq: Traceless f(R)}, we get two algebraic equations for the two potentials, namely
\begin{align}
0 &= - \kappa \bar{\rho}_m \delta_m + A_4 \frac{k^2}{a^2} \Phi + A_6 \frac{k^2}{a^2} \Psi, \label{Eq: Std A Eq} \\
0 &= \( D_4 + D_5 \frac{k^2}{a^2} \) \Phi + \( D_6 + D_7 \frac{k^2}{a^2} \) \Psi, \label{Eq: Std D Eq} 
\end{align}
from which we can solve directly for $\Phi$ and $\Psi$ in terms of $\delta_m$. The solution is
\begin{align}
\frac{k^2}{a^2} \Phi &= - \frac{\(D_6 + D_7 \frac{k^2}{a^2}\) \kappa \bar{\rho}_m \delta_m}{\(A_6 D_5 -A_4 D_7 \) \frac{k^2}{a^2} + \(A_6 D_4 - A_4 D_6 \)}, \\
\frac{k^2}{a^2} \Psi &= \frac{\(D_4 + D_5 \frac{k^2}{a^2}\) \kappa \bar{\rho}_m \delta_m}{\(A_6 D_5 - A_4 D_7 \) \frac{k^2}{a^2} + \(A_6 D_4 - A_4 D_6 \)}.
\end{align}
Note that in Eq.~\eqref{Eq: Std D Eq}, we do not neglect $D_4\Phi$ and $D_6 \Psi$. The reason is that these terms are associated to the mass of the scalaron $m_\varphi$, which can be greater than $H$ for some $f(R)$ models \cite{Tsujikawa:2007gd}. Replacing the coefficients (see Appendix \ref{App: Linear Field Equations in $f(R)$}) in the last relations we get
\begin{align}
\frac{k^2}{a^2}\Phi &= \frac{F -12 F_R ( \dot{H} + 2 H^2 )+ 2\frac{k^2}{a^2} F_R}{2 F^2 - 12 F F_R ( \dot{H} + 2 H^2 ) + 6 \frac{k^2}{a^2} F F_R} \kappa \bar{\rho}_m \delta_m, \\
\frac{k^2}{a^2} \Psi &= - \frac{F + 4\frac{k^2}{a^2} F_R}{2 F^2 - 12 F F_R ( \dot{H} + 2 H^2 ) + 6 \frac{k^2}{a^2} F F_R} \kappa \bar{\rho}_m \delta_m.
\end{align}
However, when neglecting $H$ and its derivatives, we get the following well-known expressions
\begin{equation}
\label{Eq: Standard Potentials}
\frac{k^2}{a^2} \Phi = \frac{F + 2\frac{k^2}{a^2} F_R}{2 F^2 + 6 \frac{k^2}{a^2} F F_R} \kappa \bar{\rho}_m \delta_m, \qquad \frac{k^2}{a^2} \Psi = - \frac{F + 4\frac{k^2}{a^2} F_R}{2 F^2 + 6 \frac{k^2}{a^2} F F_R} \kappa \bar{\rho}_m \delta_m.
\end{equation}
Applying the same procedure to the DE perturbations in Eqs.~\eqref{Eq: deltaDE f(R)}-\eqref{Eq: stress f(R)} we get
\begin{equation}
\label{Eq: std dDE}
\delta_\text{DE} = \frac{(1 - F)F + (2 - 3F) \frac{k^2}{a^2} F_R}{F(F + 3 \frac{k^2}{a^2} F_R)} \frac{\bar{\rho}_m}{\bar{\rho}_\text{DE}} \delta_m,
\end{equation}
\begin{equation}
\label{Eq: std dPDE}
\frac{\delta P_\text{DE}}{\bar{\rho}_\text{DE}} = \frac{1}{3F} \frac{2 \frac{k^4}{a^4} F_R + 15 \frac{k^2}{a^2} F_R \ddot{F} + 3 F \ddot{F}}{3 \frac{k^2}{a^2} F_R + F} \frac{\bar{\rho}_m}{\bar{\rho}_\text{DE}} \delta_m,
\end{equation}
\begin{equation}
\label{Eq: std VDE}
V_\text{DE} = \frac{a \dot{F}}{2F} \frac{F + 6 \frac{k^2}{a^2} F_R}{3 \frac{k^2}{a^2} F_R + F} \frac{\bar{\rho}_m}{\bar{\rho}_\text{DE}} \delta_m.
\end{equation}
\begin{equation}
\label{Eq: std Stress}
\pi_\text{DE} = \frac{\frac{k^2}{a^2}F_R}{F^2 + 3 \frac{k^2}{a^2}F F_R}\frac{\bar{\rho}_m}{\bar{\rho}_\text{DE}} \delta_m,
\end{equation}
which are the usual results found in the literature \cite{Arjona:2018jhh, Arjona:2019rfn, Cardona:2022lcz, Nesseris:2022hhc}. We will refer to the steps performed to obtain the last expressions as the ``standard QSA-SHA procedure'', while the relations for the perturbations, namely Eqs.~\eqref{Eq: Standard Potentials}-\eqref{Eq: std Stress}, will be called as the ``standard QSA-SHA results''.

We want to emphasize on one of the key steps in obtaining the above expressions. After replacing the coefficients in Appendices \ref{App: Linear Field Equations in $f(R)$} and \ref{App: Dark Energy Perturbations in $f(R)$}, we also neglected $H$ and its derivatives by assuming that these terms are subdominant under the SHA. Nonetheless, noting that $\dot{F} = F_R \dot{R}$ and $R = 6 (\dot{H} + 2 H^2)$, if we drop terms of the form $H\times$perturbation, we should naively conclude that $\dot{F} \delta_m \approx 0$, and thus the pressure perturbation and the scalar velocity would take the following form
\begin{equation}
\frac{\delta P_\text{DE}}{\bar{\rho}_\text{DE}} \approx \frac{1}{3F} \frac{2 \frac{k^2}{a^2} F_R}{3 \frac{k^2}{a^2} F_R + F} \frac{\bar{\rho}_m}{\bar{\rho}_\text{DE}} \delta_m, \qquad V_\text{DE} \approx 0.
\end{equation}
As shown in several works (see Ref. \cite{Arjona:2018jhh} for example), $V_\text{DE}$ is small at late times but certainly not zero. Indeed, as we will see in the following section, these last expressions correspond to the zero-order solution in our new parameterization of the QSA-SHA.

%%%%%%%%%%%%%%%%%%%%%%%%%%%%%%%%%%%%%%%%%%%%%%%%%%%%%%%%%%%%
\section{Tracking the Validity of the QSA and the SHA}
\label{Sec: Tracking the Accuracy of the QSA and the SHA}
%%%%%%%%%%%%%%%%%%%%%%%%%%%%%%%%%%%%%%%%%%%%%%%%%%%%%%%%%%%%

Following the discussion in the last section, there are two main assumptions behind the QSA and the SHA: the potentials do not depend on time and $k \gg a H$. These allowed us to eliminate the time derivatives of the potentials and drop some terms of the form $H\times$perturbation in the effective DE perturbations. Then, in order to know the regime where these approximations can be safely applied, we need to track the relevance of each time-derivative term in the perturbation equations. We do so by introducing the dimension-less parameters:
\begin{equation}
\label{Eq: SHA parameters}
\ve \equiv \frac{a H}{k}, \qquad \delta \equiv \frac{\dot{\ve}}{\ve H}, \qquad \xi \equiv \frac{\ddot{\ve}}{\ve H^2}, \qquad  \chi \equiv \frac{\dddot{\ve}}{\ve H^3},
\end{equation}
and
\begin{equation}
\label{Eq: QSA Parameters}
\ve_\Phi \equiv \frac{\dot{\Phi}}{\Phi H}, \qquad \ve_\Psi \equiv \frac{\dot{\Psi}}{\Psi H}, \qquad \chi_\Phi \equiv \frac{\dot{\ve}_\Phi}{\ve_\Phi H}, \qquad \chi_\Psi \equiv \frac{\dot{\ve}_\Psi}{\ve_\Psi H}.
\end{equation}
Hence, $k \gg aH$ is translated to $\ve \ll 1$, and $\dot{\Phi} \sim \dot{\Psi} \approx 0$ is translated to $\ve_\Phi \sim \ve_\Psi \ll 1$. We refer to the parameters in Eqs.~\eqref{Eq: SHA parameters} and \eqref{Eq: QSA Parameters} as ``SHA parameters'' and ``QSA parameters'', respectively.\footnote{We would like to mention that the QSA parameters are inspired by the slow-roll approximation used in the background analysis of inflationary models.}

We want to comment on some previous considerations before implementing the parameterization into the equations in Sec.~\ref{Sec: f(R) Theories} describing $f(R)$ theories. First of all, in order to track all the terms proportional to $H$ or its derivatives, we replace time derivatives of $F$ in favor of $H$, i.e.,
\begin{equation}
 \dot{F} = 6 F_R (\ddot{H} + 4 H\dot{H}), \quad \ddot{F} = 6 F_R (\dddot{H} + 4 H \ddot{H} + 4 \dot{H}^2) + 36 F_{RR} (\ddot{H} + 4 H \dot{H})^2.
\end{equation}
In addition, apart of the parameter $\ve$, all the other SHA parameters are not necessarily small. For example, during matter domination we have $H = H_0 \sqrt{\Omega_{m0}} a^{-3/2}$ and
\begin{equation}
\label{Eq: SHA in MD}
\delta = -1/2, \quad \xi = 1, \quad \chi = - 7 / 2, \qquad \text{(in matter domination)}
\end{equation}
where $\Omega_{m0}$ is the matter density parameter today, and $H_0$ is the Hubble constant. 

Re-writing Eqs.~\eqref{Eq: 00 f(R)} and \eqref{Eq: Traceless f(R)} using the new parameters, we obtain the following two ``algebraic'' equations to solve for $\Phi$ and $\Psi$
\begin{align}
0 &= - \kappa \bar{\rho}_m \delta_m + \mc{A}_1 \frac{k^2}{a^2} \Phi + \mc{A}_2 \frac{k^2 \ve^2}{a^2} \Phi + \mc{A}_3 \frac{k^4 \ve^4}{a^4} \Phi + \mc{A}_4 \frac{k^2}{a^2} \Psi + \mc{A}_5 \frac{k^2 \ve^2}{a^2} \Psi + \mc{A}_6 \frac{k^4 \ve^4}{a^4} \Psi, \label{Eq: New 00 eq} \\
0 &= \( \mc{D}_1 + \mc{D}_2 \frac{k^2}{a^2} \) \Phi + \mc{D}_3 \frac{k^2 \ve^2}{a^2} \Phi + \( \mc{D}_4 + \mc{D}_5 \frac{k^2}{a^2}\) \Psi + \mc{D}_6 \frac{k^2 \ve^2}{a^2} \Psi, \label{Eq: New traceless eq}
\end{align}
where the coefficients $\mc{A}_i$ and $\mc{D}_i$ are found in Appendix \ref{App: Linear Field Equations in the New Parametrization}. Note that we wrote the above equations factorizing in terms of $\ve$ to facilitate the measurement of the relevance of each term. Now, solving for $\Phi$ and $\Psi$, expanding in powers of $\ve$ up to second-order, replacing the coefficients $\mc{A}_i$ and $\mc{D}_i$, and considering only linear terms in the QSA parameters we get%\footnote{The full expressions for $\Phi$ and $\Psi$ can be found in Appendix \ref{App: Full Expression for the Potentials}.}
\begin{align}
\frac{k^2}{a^2} \Phi &= \frac{F + 2\frac{k^2}{a^2} F_R}{2 F^2 + 6 \frac{k^2}{a^2} F F_R} \kappa \bar{\rho}_m \delta_m + \frac{3 \kappa \bar{\rho}_m \delta_m \ve^2}{4 F \( F + 3 \frac{k^2}{a^2} F_R \)^2} \Big\{ F^2 (2 + \ve_\Phi + \ve_\Psi) \label{Eq: 2º Phi} \\
 &+ 2 \frac{k^2}{a^2} F F_R [ \delta(1 + \ve_\Phi) + 4(2 + \ve_\Psi) + 5 \ve_\Phi ]+ 4 \frac{k^4}{a^4} F_R^2[ \delta (3 + \ve_\Phi) + 4 (2 + \ve_\Psi) + 4 \ve_\Phi] \Big\}, \nonumber \\
\frac{k^2}{a^2} \Psi &= - \frac{F + 4\frac{k^2}{a^2} F_R}{2 F^2 + 6 \frac{k^2}{a^2} F F_R} \kappa \bar{\rho}_m \delta_m + \frac{3 \kappa \bar{\rho}_m \delta_m \ve^2}{4 F \( F + 3 \frac{k^2}{a^2} F_R \)^2} \Big\{ F^2 (2 + \ve_\Phi + \ve_\Psi) \label{Eq: 2º Psi} \\
 &- 2 \frac{k^2}{a^2} F F_R [ \delta(3 + \ve_\Phi) - 3 (2 + \ve_\Psi)] - 4 \frac{k^4}{a^4} F_R^2[ \delta (6 + \ve_\Phi) - 2 (2 + \ve_\Psi) + \ve_\Phi] \Big\}. \nonumber
\end{align}
We recognize the standard expressions for the potentials [see Eq.~\eqref{Eq: Standard Potentials}] in the first term on the right-hand side of the last expressions. Hence, we can identify the standard results for the potentials as the solution for the \textit{deepest} modes in the Hubble radius or, equivalently, the smallest scales where $\ve \sim 0$. 

Indeed, equations \eqref{Eq: New 00 eq} and \eqref{Eq: New traceless eq} are not ``algebraic'' equations for $\Phi$ and $\Psi$, since the QSA parameters $\ve_\Phi$ and $\ve_\Psi$ depends on the dynamics of the potentials. Therefore, Eqs.~\eqref{Eq: 2º Phi} and \eqref{Eq: 2º Psi} are coupled differential equations for $\Phi$ and $\Psi$. However, note that the QSA parameters are always coupled to SHA parameters, and thus terms of the form $\varepsilon_\Phi \times \varepsilon^2$ are third-order terms in the new QSA-SHA expansion. We will consider only second-order terms, and thus we neglect the QSA parameters, i.e., we will assume that $\ve_\Phi = \ve_\Psi = 0$. As we will see, these approximations, zero-order in QSA parameters and second-order in SHA parameters, are sufficiently accurate when compare against full numerical results for large modes. Under this assumptions, the expressions for the potentials are reduced to
\begin{align}
\frac{k^2}{a^2} \Phi &= \frac{F + 2\frac{k^2}{a^2} F_R}{2 F^2 + 6 \frac{k^2}{a^2} F F_R} \kappa \bar{\rho}_m \delta_m \label{Eq: No QSA Phi} \\
 & + \frac{3 \kappa \bar{\rho}_m \delta_m \ve^2}{2 F \( F + 3 \frac{k^2}{a^2} F_R \)^2} \Big\{ F^2 + \frac{k^2}{a^2} ( \delta + 8 )F F_R + 2 \frac{k^4}{a^4}( 3 \delta + 8 ) F_R^2 \Big\}, \nonumber \\
\frac{k^2}{a^2} \Psi &= - \frac{F + 4\frac{k^2}{a^2} F_R}{2 F^2 + 6 \frac{k^2}{a^2} F F_R} \kappa \bar{\rho}_m \delta_m \label{Eq: No QSA Psi} \\
 &+ \frac{3 \kappa \bar{\rho}_m \delta_m \ve^2}{2 F \( F + 3 \frac{k^2}{a^2} F_R \)^2} \Big\{ F^2 - \frac{k^2}{a^2} ( 3\delta - 6) F F_R - 2 \frac{k^4}{a^4} ( 6\delta - 4) F_R^2 \Big\}. \nonumber
\end{align}
Now, applying the new parameterization to the DE perturbations quantities in Eqs.~\eqref{Eq: deltaDE f(R)}-\eqref{Eq: stress f(R)}, and then inserting these potentials we get:
\begin{align}
\delta_\text{DE} &= \frac{(1 - F)F + \frac{k^2}{a^2} (2 - 3F) F_R}{F(F + 3 \frac{k^2}{a^2} F_R)} \frac{\bar{\rho}_m}{\bar{\rho}_\text{DE}} \delta_m \label{Eq: New deltaDE} \\
 &- \frac{3 \frac{k^2}{a^2} F_R \ve^2}{F \(F + 3 \frac{k^2}{a^2} F_R \)^2} \Big\{ ( \delta + 1) F + 2 \frac{k^2}{a^2} ( 3\delta + 2) F_R \Big\} \frac{\bar{\rho}_m}{\bar{\rho}_\text{DE}} \delta_m, \nonumber \\
\frac{\delta P_\text{DE}}{\bar{\rho}_\text{DE}} &= \frac{1}{3F} \frac{2 \frac{k^2}{a^2} F_R}{3 \frac{k^2}{a^2} F_R + F} \frac{\bar{\rho}_m}{\bar{\rho}_\text{DE}} \delta_m + \frac{3 \frac{\bar{\rho}_m}{\bar{\rho}_\text{DE}} \delta_m \ve^2}{F \(F + 3 \frac{k^2}{a^2} F_R \)^2} \Big\{ F^2 (F - 1)(1 + 2 \delta) \label{Eq: New deltaPDE} \\
 & + \frac{k^2}{a^2}(5F -10\delta + 13 F \delta - 5) F F_R + \frac{k^4}{a^4}(6F - 6 \delta + 21 F \delta - 4) F_R^2 \Big\}, \nonumber \\
V_\text{DE} &= \frac{a}{F \(F + 3 \frac{k^2}{a^2} F_R \)} \Big\{ F (F - 1) + \frac{k^2}{a^2} (3F - 4) F_R \Big\} \frac{\bar{\rho}_m}{\bar{\rho}_\text{DE}} \delta_m \frac{k}{a} \ve \label{Eq: New VDE}, \\
\pi_\text{DE} &= \frac{\frac{k^2}{a^2}F_R}{F^2 + 3 \frac{k^2}{a^2}F F_R}\frac{\bar{\rho}_m}{\bar{\rho}_\text{DE}} \delta_m + \frac{3 F_R \frac{\bar{\rho}_m}{\bar{\rho}_\text{DE}} \delta_m \frac{k^2}{a^2} \ve^2}{F \(F + 3 \frac{k^2}{a^2} F_R \)^2} \Big\{ F(1 + 2 \delta) + \frac{k^2}{a^2} F_R (4 + 9 \delta) \Big\}. \label{Eq: New stress}
\end{align}
From the last expressions, we can unravel similarities and differences between the standard QSA-SHA results in Eqs.~\eqref{Eq: Standard Potentials}-\eqref{Eq: std Stress} and our expressions in Eqs.~\eqref{Eq: No QSA Phi}-\eqref{Eq: New stress}. For instance, we see that the standard $\delta_\text{DE}$ in Eq.~\eqref{Eq: std dDE} corresponds to the zero-order contribution in Eq.~\eqref{Eq: New deltaDE}. In the new parameterization, $\delta P_\text{DE}$ has contributions to second-order in $\ve$, while $V_\text{DE}$ is predominantly a first-order quantity in the $\ve$-expansion. Now, bearing in mind that
\begin{equation}
\dot{F} = 6 \frac{k^3}{a^3} (\delta + \xi - 2) F_R \ve^3, 
\end{equation}
\begin{equation}
\label{Eq: ddot F}
\ddot{F} = 6 \frac{k^4}{a^4} (\delta^2 - 8 \delta + \chi + 6) F_R \ve^4 + 36 \frac{k^6}{a^6} (\delta + \xi - 2) F_{RR} \ve^6,
\end{equation}
we note that, in the new parameterization, the standard $\delta P_\text{DE}$ in Eq.~\eqref{Eq: std dPDE} corresponds to a zero-order quantity plus fourth and sixth-order corrections, while the standard $V_\text{DE}$ in Eq.~\eqref{Eq: std VDE} is entirely a third-order quantity. 

Equations \eqref{Eq: No QSA Phi}-\eqref{Eq: New stress} are the main results of our paper. They explicitly show that potentially relevant terms have been neglected in the effective DE perturbations when the QSA and the SHA are applied in the standard way. They also point out that similar expressions obtained for other MG theories, such as Horndeski theories, neglected terms which could potentially modify their predictions. Now, in order to determine the relevance of the correction terms uncovered by our new QSA-SHA procedure, we have to know the evolution of the SHA parameters which depend on $H$ and its derivatives, i.e., they depend on the specific cosmological model. In the following section we will study the behavior of these parameters in the context of two well-known $f(R)$ models: the designer model \cite{Multamaki:2005zs, delaCruz-Dombriz:2006kob} and the Hu-Sawicki model \cite{Hu:2007pj}.

%%%%%%%%%%%%%%%%%%%%%%%%%%%%%%%%%%%%%%%%%%%%%%%
\section{Specific $f(R)$ Models}
\label{Sec: Specific Models}
%%%%%%%%%%%%%%%%%%%%%%%%%%%%%%%%%%%%%%%%%%%%%%%

%%%%%%%%%%%%%%%%%%%%%%%%%%%%
\subsection{$f$DES Model}
%%%%%%%%%%%%%%%%%%%%%%%%%%%%

As shown in Refs. \cite{Capozziello:2005ku, Nojiri:2006gh}, $f(R)$ models admit solutions which can be ``designed'' to reproduce any desirable expansion history. In particular, it is possible to find an $f(R)$ model matching the $\Lambda$CDM background but having deviations at the perturbative level. The Lagrangian of this designer model, $f$DES hereafter, reads \cite{Multamaki:2005zs} 
\begin{equation}
f(R) = R - 2 \Lambda + \alpha H_0^2 \( \frac{\Lambda}{R - 3 \Lambda} \)^{c_0} {}_2F_1 \( c_0, \frac{3}{2} + c_0, \frac{13}{6} + 2c_0, \frac{\Lambda}{R - 3 \Lambda} \),
\end{equation}
where $c_0 \equiv (\sqrt{73} - 7)/12$, $\alpha$ is a dimensionless constant, and ${}_2 F_1$ denotes the hypergeometric function. Since the background is the same as in $\Lambda$CDM, from Eqs.~\eqref{Eq: f(R) density} and \eqref{Eq: f(R) pressure} we have
\begin{equation}
\label{Eq: H fDES}
H^2 = H_0^2 \( \Omega_{m0} a^{-3} + \Omega_{\Lambda0} \), \qquad \bar{\rho}_\text{DE} = - \bar{P}_\text{DE} = 3 H_0^2 \Omega_{\Lambda0},
\end{equation}
where $\Omega_{\Lambda0}$ is the DE density parameter today. The parameter $\alpha$  can be estimated following the prescription that viable $f(R)$ models require
\begin{equation}
f_{R0} \equiv F(a = 1) - 1 \ll 1.
\end{equation}
A typical value is $f_{R0} \sim - 10^{-4}$ (see, for instance, Ref. \cite{Pogosian:2007sw}) which translates to $\alpha \approx 0.00107$.

%%%%%%%%%%%%%%%%%%%%%%%%%%%%%%%
\subsection{Hu-Sawicki Model}
%%%%%%%%%%%%%%%%%%%%%%%%%%%%%%%

\begin{figure}[t!]
\includegraphics[width = 0.52\textwidth]{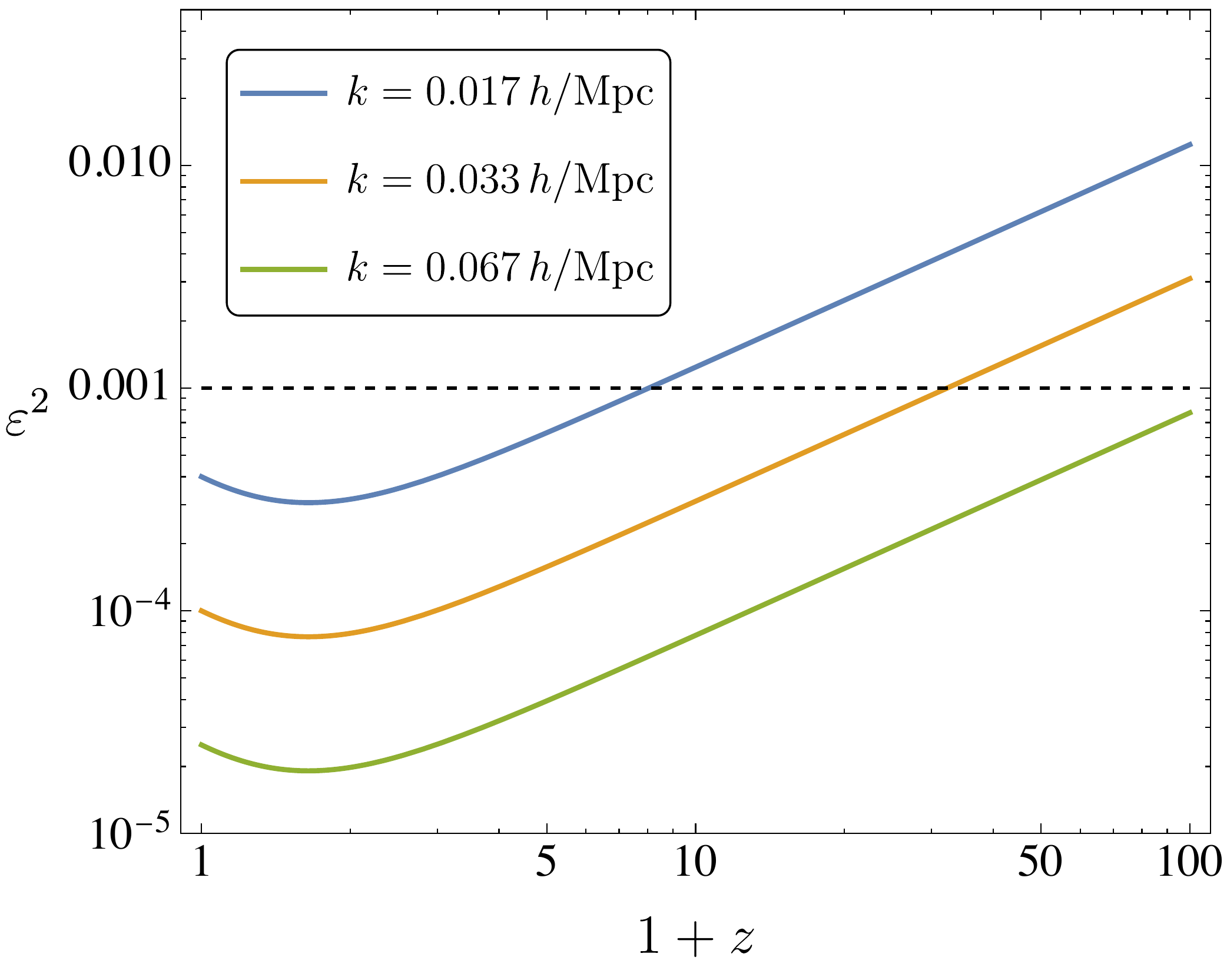}
\hfill
\includegraphics[width = 0.47\textwidth]{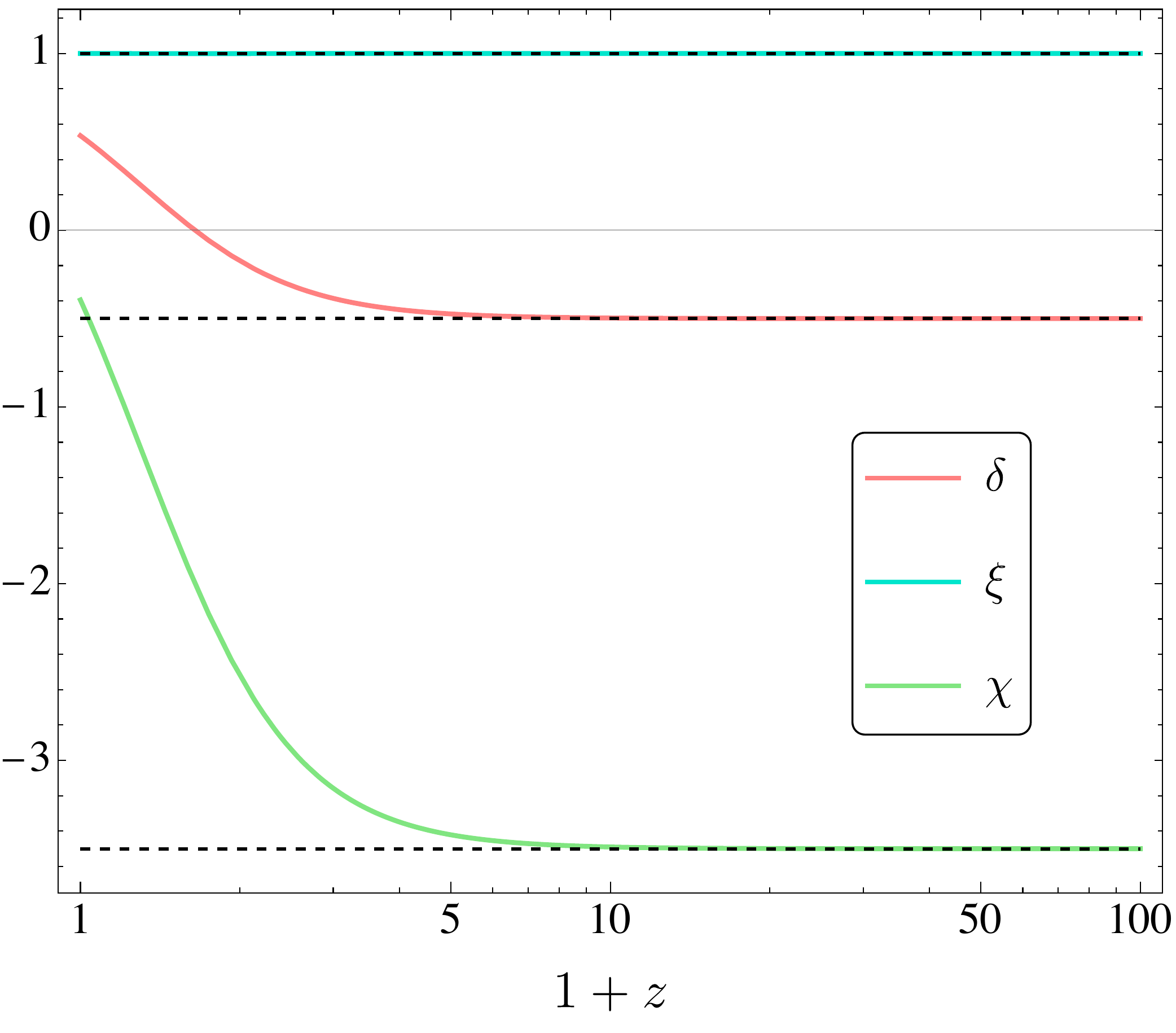}
\caption{Left: Evolution of the $\ve$ parameter in the HS model for three values of $k$. We can see that $\ve^2 \lesssim 10^{-3}$ during the whole matter dominated epoch until present days for scales $k \gtrsim 0.06 \, h/\text{Mpc}$. Right: Evolution of the SHA parameters $\delta$, $\xi$, and $\chi$ in the HS model. Note that they behave as constants during the matter dominated epoch as expected from Eqs.~\eqref{Eq: SHA in MD} (see the dashed black lines), with a raising at late-times when the contribution of DE becomes more relevant.}
\label{Fig: SHA Parameters}
\end{figure}

The Hu-Sawicki (HS) model is a well-known model constructed in such a way that observational tests, such as solar system tests, can be surpassed. Its Lagrangian is given by \cite{Hu:2007pj}
\begin{equation}
f(R) = R - m^2 \frac{c_1 (R / m^2)^n}{1 + c_2 (R / m^2)^n},
\end{equation}
where $m$, $c_1$, $c_2$, and $n$ are dimensionless constants. Through algebraic manipulation, it is possible to re-write the last function as \cite{Basilakos:2013nfa}
\begin{equation}
f(R) = R - \frac{2 \Lambda}{1 + \(\frac{b \Lambda}{R}\)^n}, \quad \Lambda = \frac{m^2 c_1}{2 c_2}, \quad b = \frac{2 c_2^{1 - 1/n}}{c_1}.
\end{equation}
Therefore, the background evolution in the HS model can be similar to $\Lambda$CDM depending on $b$ and $n$. Usually $n$ takes on positive integer values. Here, we choose $n = 1$. Assuming $f_{R0} = - 10^{-4}$, we estimate $b \approx 0.000989557$. 

In order to characterize the background evolution of the HS model, in general we would have to solve the Friedmann equations \eqref{Eq: Friedman Eqs} and the continuity equations \eqref{Eq: Continuity Eqs}. However, we can evade this issue by using the following analytical approximation to the Hubble parameter \cite{Basilakos:2013nfa}
\begin{equation}
\label{Eq: H HS}
H_\mathrm{HS}(a)^2=H_\Lambda(a)^2+b~\delta H_1(a)^2+\mathcal{O}(b^2),
\end{equation}
where the function $\delta H_1(a)$ is given in the Appendix of Ref.~\cite{Basilakos:2013nfa} and $H_\Lambda$ is the Hubble parameter for the $\Lambda$CDM model. Overall, Eq.~\eqref{Eq: H HS} is accurate to better than $\sim 10^{-5} \%$, for our choice of the parameter $b$.

Having the functional forms of $H$, Eqs.~\eqref{Eq: H fDES} and \eqref{Eq: H HS} for the $f$DES and HS models respectively, we can follow the evolution of the SHA parameters defined in Eq.~\eqref{Eq: SHA parameters}. However, as noted in Eqs.~\eqref{Eq: H fDES} and \eqref{Eq: H HS}, the HS can be regarded as a small deviation from the $\Lambda$CDM model for $b$ small enough. We verified that indeed the evolution of the SHA parameters in both models are essentially indistinguishable for the values of $\alpha$ and $b$ we chose. Then, to keep our presentation simple, in Fig.~\ref{Fig: SHA Parameters} we only plot the evolution of the SHA parameters for the HS model using as time variable the redshift $z$, which is related to the scale factor through the relation $a = 1/(1 + z)$. In the left-panel of this figure, we see that $\ve^2$ can be safely considered as ``small'' (i.e., $\ve^2 \lesssim 10^{-3}$) for modes\footnote{In completely theoretical studies, $k$ is typically given in units of $H_0$. Therefore, $k \approx 0.067 \, h/\text{Mpc}$ corresponds to $k \approx 200 H_0$.} $k \gtrsim 0.06 \, h/\text{Mpc}$. In the right panel we see that at earlier times the other SHA parameters behave as expected during the matter dominated epoch [see Eq.~\eqref{Eq: SHA in MD}] and grow when DE plays a more relevant role in the cosmic budget. Therefore, we say that the ``safety region'' where the SHA can be applied corresponds to the scales $0.067 \, h/\text{Mpc} \lesssim k \lesssim 0.2 \, h / \text{Mpc}$. The upper bound of this window scale comes from the fact that this is roughly the scale above which nonlinearities at present days cannot be ignored in the concordance model\footnote{The scale of nonlinearities $k_\text{NL}$ is defined as the scale for which the ``dimensionless'' linear matter spectrum is equal to 1. In the case of $\Lambda$CDM, it is given by $k_\text{NL} \approx 0.25 \, h/\text{Mpc}$ today \cite{dodelson2020}.} \cite{dodelson2020}. This scale is also close to the maximum comoving wave number associated with the galaxy power spectrum measured by the Sloan Digital Sky Survey (SDSS) without entering the non-linear regime  \cite{SDSS:2003tbn, DeFelice:2016uil}.

We would like to point out that the $f$DES model and the HS model have a background evolution either identical or similar to that of $\Lambda$CDM, and that $f(R)$ models with different background evolution could yield to different results, in principle. Nonetheless, as clarified in Ref. \cite{Amendola:2006we}, $f(R)$ models with expansion histories similar to that of $\Lambda$CDM are cosmologically viable, while more generic models have in general to meet a number of requirements, making the model-building process more contrived \cite{Battye:2017ysh}.

%%%%%%%%%%%%%%%%%%%%%%%%%%%%%%%%%%%%%%%%%%%%%%%%%%%%%%%%%%
\section{Numerical Solution of the Evolution Equations}
\label{Sec: Numerical Solution}
%%%%%%%%%%%%%%%%%%%%%%%%%%%%%%%%%%%%%%%%%%%%%%%%%%%%%%%%%%

In this section we numerically solve the equations for matter perturbations in Eqs.~\eqref{Eq: dm Eq} and \eqref{Eq: Vm Eq} assuming that $\Phi$ and $\Psi$ are given by: $i)$ the standard QSA and SHA expressions in Eqs.~\eqref{Eq: Standard Potentials}, and $ii)$ the new parameterized expressions in Eqs.~\eqref{Eq: No QSA Phi} and \eqref{Eq: No QSA Psi}. Then, we use these solutions to find the evolution of the DE perturbations given by the standard procedure in Eqs.~\eqref{Eq: std dDE} and \eqref{Eq: std VDE}, and the second-order SHA expressions in Eqs.~\eqref{Eq: New deltaDE}-\eqref{Eq: New VDE}. We compare these results against full numerical solutions where no approximations are assumed. This will allow us to assess the relevance of the $O(\varepsilon^2)$ correction terms to the standard results.

%%%%%%%%%%%%%%%%%%%%%%%%%%%%%%%%%%%%%%%%
\subsection{Full Numerical Solution}
%%%%%%%%%%%%%%%%%%%%%%%%%%%%%%%%%%%%%%%%

It is desirable to solve the full set of linear equations in Eqs.~\eqref{Eq: 00 f(R)}-\eqref{Eq: Traceless f(R)} complemented with the continuity equations in Eqs.~\eqref{Eq: dm Eq} and \eqref{Eq: Vm Eq}. However, this set of equations is highly unstable, and typical numerical methods do not yield to reliable results. In Ref. \cite{Pogosian:2007sw}, this problem was solved by rendering the perturbation equations in a more treatable numerical way which we describe in the following.

Multiplying the ``Time-Space'' equation \eqref{Eq: 0i f(R)} by $-3H$ and using the ``Time-Time'' equation \eqref{Eq: 00 f(R)} we get the following Poisson-like equation
\begin{equation}
\label{Eq: Poisson Eq}
\frac{k^2}{a^2}(\Phi - \Psi) F = \kappa \bar{\rho}_m \Delta_m - 3 \dot{F} \dot{\Phi} - 3 F \dot{H} \Phi + (3 H \dot{F} - 3 F \dot{H}) \Psi,
\end{equation}
where we have defined the matter comoving density perturbation
\begin{equation}
\Delta_m = \delta_m + \frac{3 a H V_m}{k^2}.
\end{equation}
Now, we define the variables
\begin{equation}
\Phi_+ = \frac{1}{2}(\Phi - \Psi), \qquad \zeta = - F (\Phi + \Psi).
\end{equation}
The evolution equations for these new variables can be obtained from the ``Time-Space'' equation \eqref{Eq: 0i f(R)} and the Poisson-like equation \eqref{Eq: Poisson Eq}. We get
\begin{align}
\Phi\text{'}_+ &= - \frac{a \kappa \bar{\rho}_m V_m}{2 F H k^2} - \( 1 + \frac{F\text{'}}{2 F} \) \Phi_+ - \frac{3 F\text{'}}{4 F^2} \zeta, \\
\zeta\text{'} &= - \frac{2 \kappa \bar{\rho}_m \Delta_m}{3H^2} \frac{F}{F\text{'}} + \( 1 + \frac{F\text{'}}{F} - 2 \frac{F}{F\text{'}} \frac{H\text{'}}{H} \) \zeta + 2 F \Phi\text{'}_+ + 2F \( 1 + \frac{2}{3}\frac{k^2}{a^2 H^2} \frac{F^2}{F\text{'}} \) \Phi_+,
\end{align}
where a quote $\text{'}$ means derivative with respect to the number of $e$-folds, which is related to the time-derivative by $\text{d} N \equiv H \text{d}t$. These equations are complemented with the equations for $V_m$ and $\delta_m$ in Eqs.~\eqref{Eq: Vm Eq} and \eqref{Eq: dm Eq}, having in mind that the relation between scale factor-derivatives and $e$-folds-derivatives is $\text{d} a \equiv a \text{d} N$. We assume that there are no deviations from GR at early times, then the initial conditions are
\begin{equation}
\Phi_{+, i} = 1, \qquad \zeta_i = 0, \qquad V_{m, i} = - \frac{2}{3} \frac{k^2}{a_i H_i} \Phi_{+, i}, \qquad \Delta_{m, i} = \frac{2 k^2}{3 a_i^2 H_i^2} \Phi_{+, i},
\end{equation}
where the expressions for $V_{m, i}$ and $\Delta_{m, i}$ correspond to the simplification of the ``Time-Space'' equation \eqref{Eq: 0i f(R)} and the Poisson equation \eqref{Eq: Poisson Eq} in matter dominance, i.e., taking $H^2 = H_0^2 \Omega_{m 0} a^{-3}$, and assuming that early modifications to GR are negligible. For $H_i$, we choose $a_i = 10^{-3}$ ($z \approx 1000$) ensuring initial conditions well within the matter epoch, right after decoupling.

%%%%%%%%%%%%%%%%%%%%%%%%%%%%%%%%%%
\subsection{Comparing Results}
%%%%%%%%%%%%%%%%%%%%%%%%%%%%%%%%%%

\begin{figure}[t!]
\includegraphics[width = 0.495\textwidth]{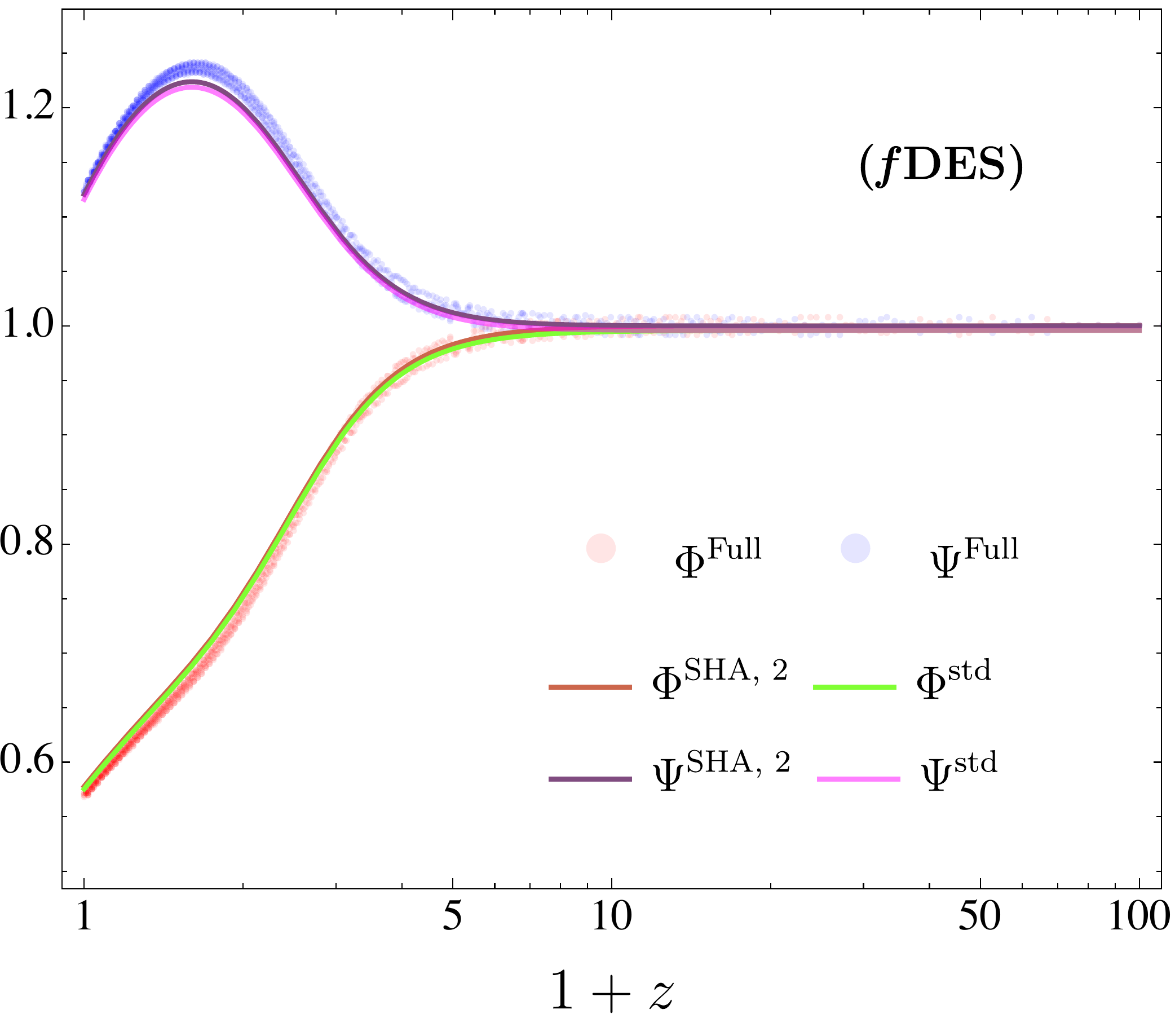}
\hfill
\includegraphics[width = 0.495\textwidth]{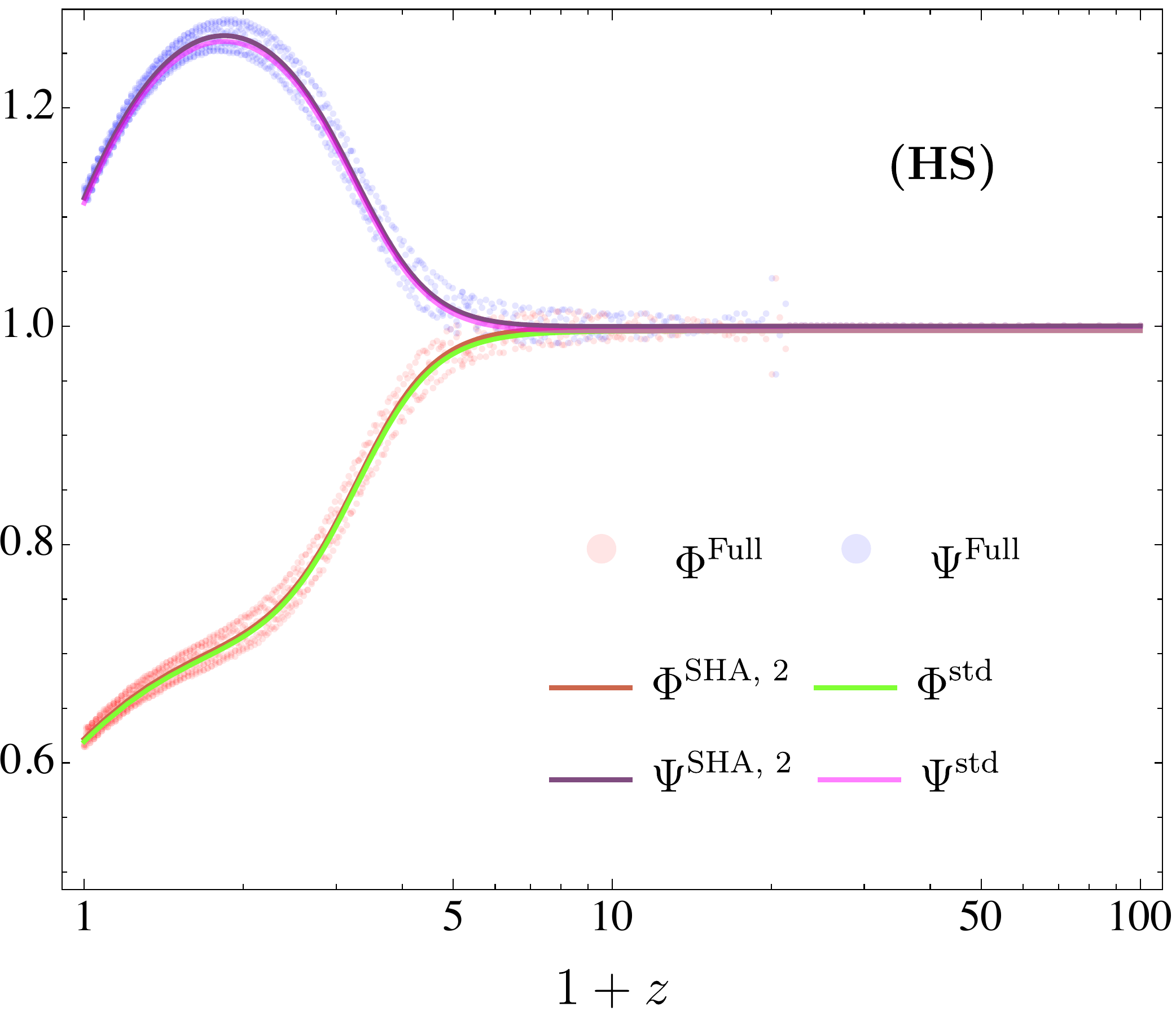}
\caption{Comparison of the potentials $\Phi$ and $\Psi$ obtained from the full numerical solution (Full), from the standard QSA-SHA procedure (std), and from our new approach (SHA, 2). Left panel corresponds to the results for the $f$DES model, while the right panel corresponds to the results for the HS model. For the plots we used $k = 0.1 \, h/\text{Mpc}$.}
\label{Fig: Potentials}
\end{figure}

In this section, we compare the full numerical solutions, which are obtained using the approach described in the last section, the standard QSA-SHA results (see Sec.~\ref{Sec: f(R) Theories}), and the results from our new parameterization described in Sec.~\ref{Sec: Tracking the Accuracy of the QSA and the SHA}, for the two models considered, namely, the $f$DES model and the HS model.

%%%%%%%%%%%%%%%%%%%%%%%%%%%%%%%%%%%%%%%%%%
\subsubsection{Gravitational Potentials}
%%%%%%%%%%%%%%%%%%%%%%%%%%%%%%%%%%%%%%%%%%

In order to solve the evolution equations for $\delta_m$ and $V_m$ in Eqs.~\eqref{Eq: dm Eq} and \eqref{Eq: Vm Eq}, we replace the analytic expressions for the potentials $\Phi$ and $\Psi$ from either the standard procedure in Eqs.~\eqref{Eq: Standard Potentials} or our new expressions in Eqs.~\eqref{Eq: No QSA Phi} and \eqref{Eq: No QSA Psi}, using the initial conditions \cite{Sapone:2009mb}:
\begin{equation}
\delta_{m, i} = \delta_i\, a_i \(1 + 3 \frac{a_i^2 H^2(a_i)}{k^2} \), \quad V_{m, i} = - \delta_i\, H_0\, \Omega_{m 0}\, a_i^{1/2}.
\end{equation}
where the overall factor $\delta_i$ is set to unity. Then, knowing $\delta_m$ it is possible to track the evolution of the potentials and, consequently, of the DE perturbations.

As it can be seen in Fig.~\ref{Fig: Potentials}, the full solutions for the potentials for the two models undergo very rapid oscillations with small amplitudes, i.e., they remain well-behaved. This oscillatory behavior was also pointed out in Refs. \cite{Starobinsky:2007hu, Pogosian:2007sw}, where it was attributed to the evolution of $\delta R$, this being a generic feature of $f(R)$ models. In both plots, left-panel for the $f$DES model and right panel for the HS model, we can see that the standard QSA-SHA procedure and our new approach give accurate approximations to the full numerical solutions. On these scales, $k = 0.1 \, h/\text{Mpc}$, $\varepsilon^2 \ll 1$ and thus the second-order correction terms to the standard QSA-SHA are not compelling.

\begin{figure}[t!]
\includegraphics[width = 0.495\textwidth]{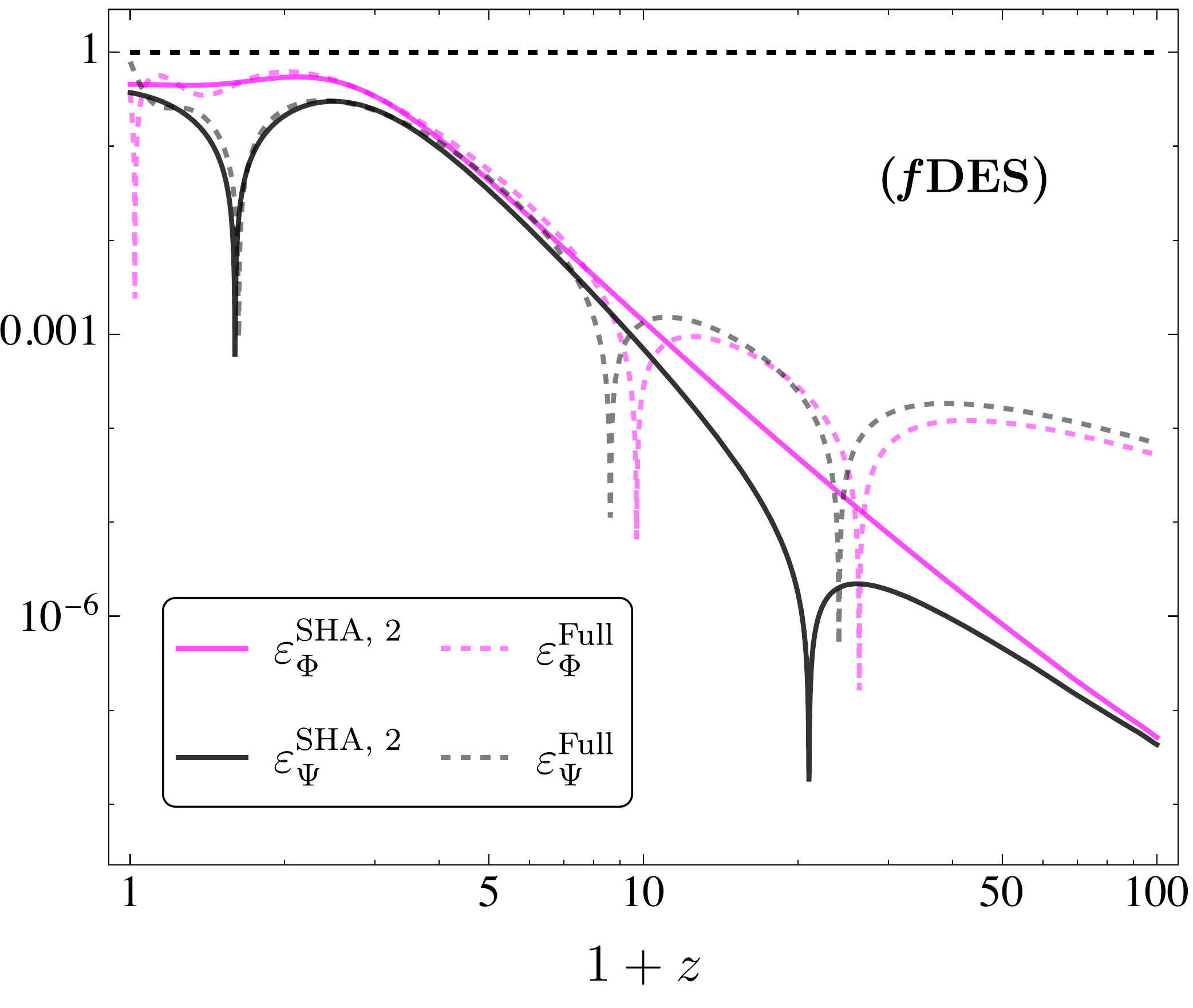}
\hfill
\includegraphics[width = 0.495\textwidth]{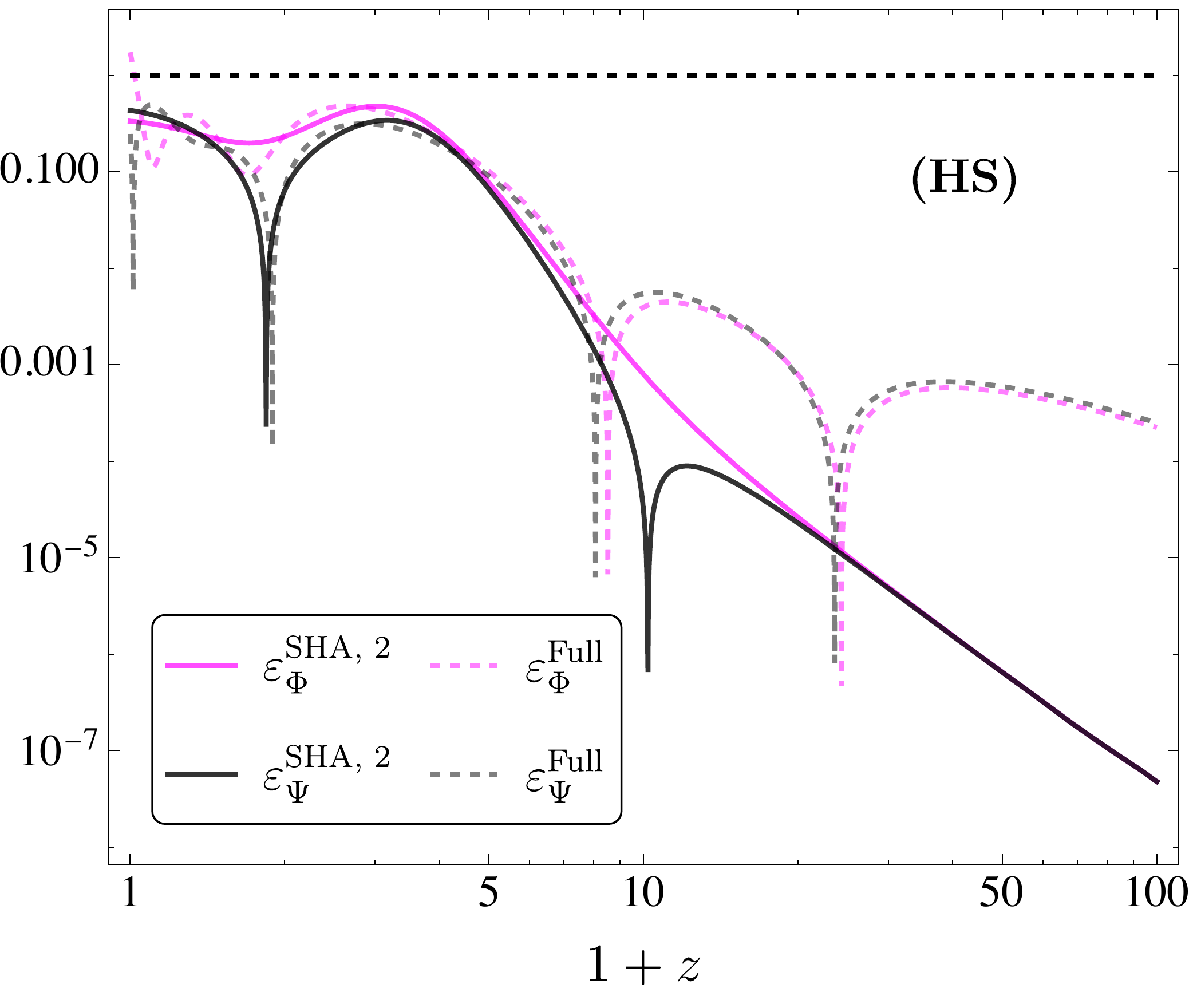}
\caption{Evolution of the QSA parameters $\varepsilon_\Phi$ and $\varepsilon_\Psi$ defined in Eqs.~\eqref{Eq: QSA Parameters}. For both models ($f$DES on the left panel and HS on the right panel), these parameters are small during the matter dominated epoch and significantly increase around the matter-DE transition. However, they are always suppressed by $\varepsilon^2$ and thus third-order terms, ($\varepsilon_\Phi \times \varepsilon^2$) for example, are negligible.}
\label{Fig: QSA Parameters}
\end{figure}

From the results for the potentials, it is possible to compute the QSA parameters $\ve_\Phi$ and $\ve_\Psi$ in order to determine the time and scale validity regimes of the QSA; as we did for the SHA in Sec.~\ref{Sec: Specific Models}. As depicted in Fig.~\ref{Fig: QSA Parameters}, these parameters are very small ($\ll 10^{-3}$) during the whole matter dominated epoch, with a significant increasing at later times. However, as mentioned before, these parameters are always coupled to $\varepsilon^2$ in the expression for the potentials in Eqs.~\eqref{Eq: 2º Phi} and \eqref{Eq: 2º Psi}. Therefore, they are subdominant in comparison with purely $\varepsilon^2$ terms. We then conclude that in the case of $f(R)$ models under consideration, the assumption that the contributions of the time-derivatives of the potentials to the cosmological dynamics are negligible can be safely applied for the scales when the SHA is also applicable. 

%%%%%%%%%%%%%%%%%%%%%%%%%%%%%%%%%%%%%%%%%%%%
\subsubsection{Dark Energy Perturbations}
%%%%%%%%%%%%%%%%%%%%%%%%%%%%%%%%%%%%%%%%%%%%

\begin{figure}[t!]
\includegraphics[width = 0.495\textwidth]{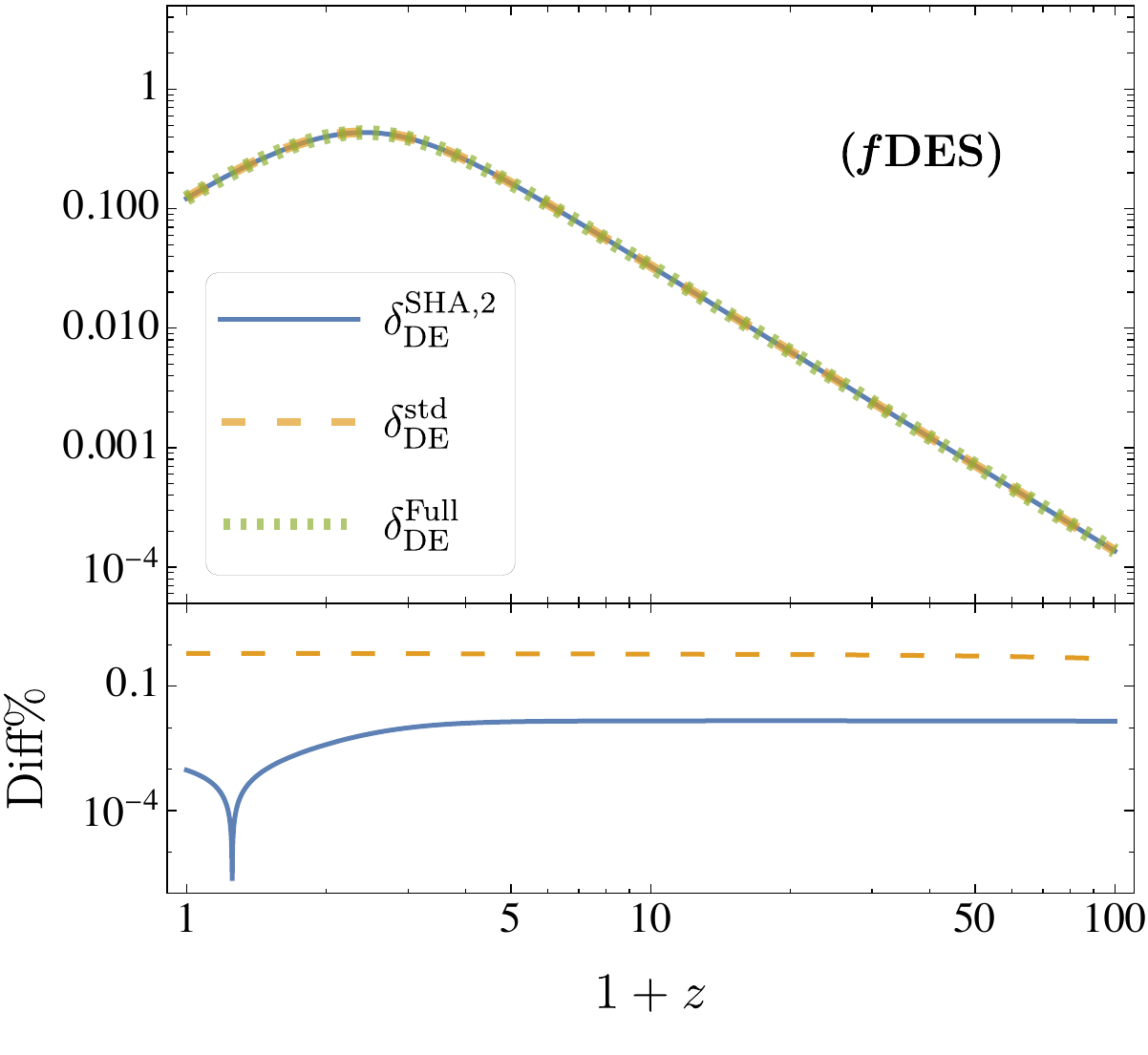}
\hfill
\includegraphics[width = 0.495\textwidth]{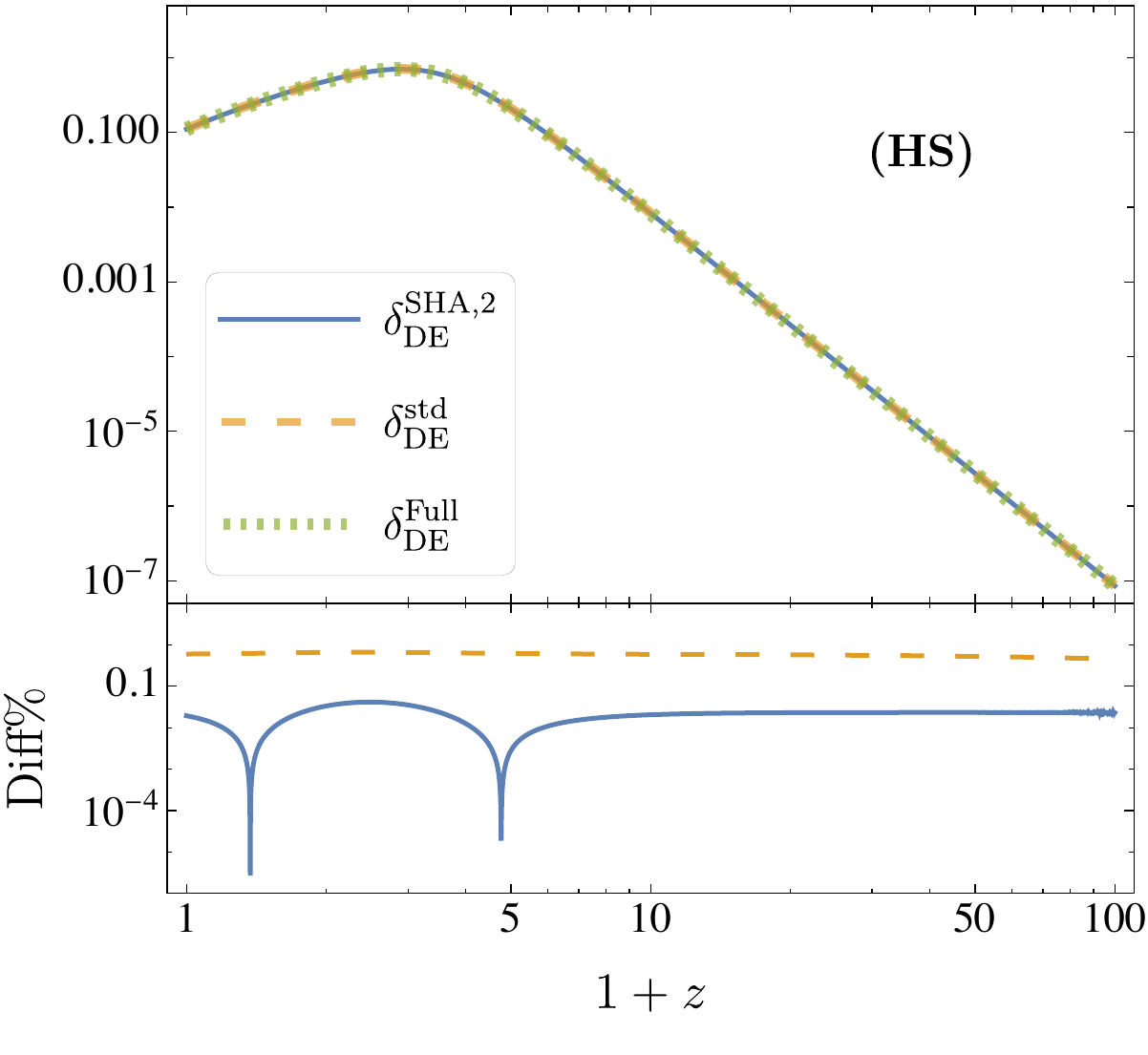}
\caption{Comparison of the DE contrast $\delta_\text{DE}$ obtained from full numerical solution, the standard QSA-SHA approach, and our new parameterization. For both models ($f$DES on the left panel and HS on the right panel), both descriptions are accurate at sub-percent level. However, it is noteworthy that our approach improve over the standard result by around two orders of magnitude.}
\label{Fig: deltas}
\end{figure}

\begin{figure}[t!]
\includegraphics[width = 0.495\textwidth]{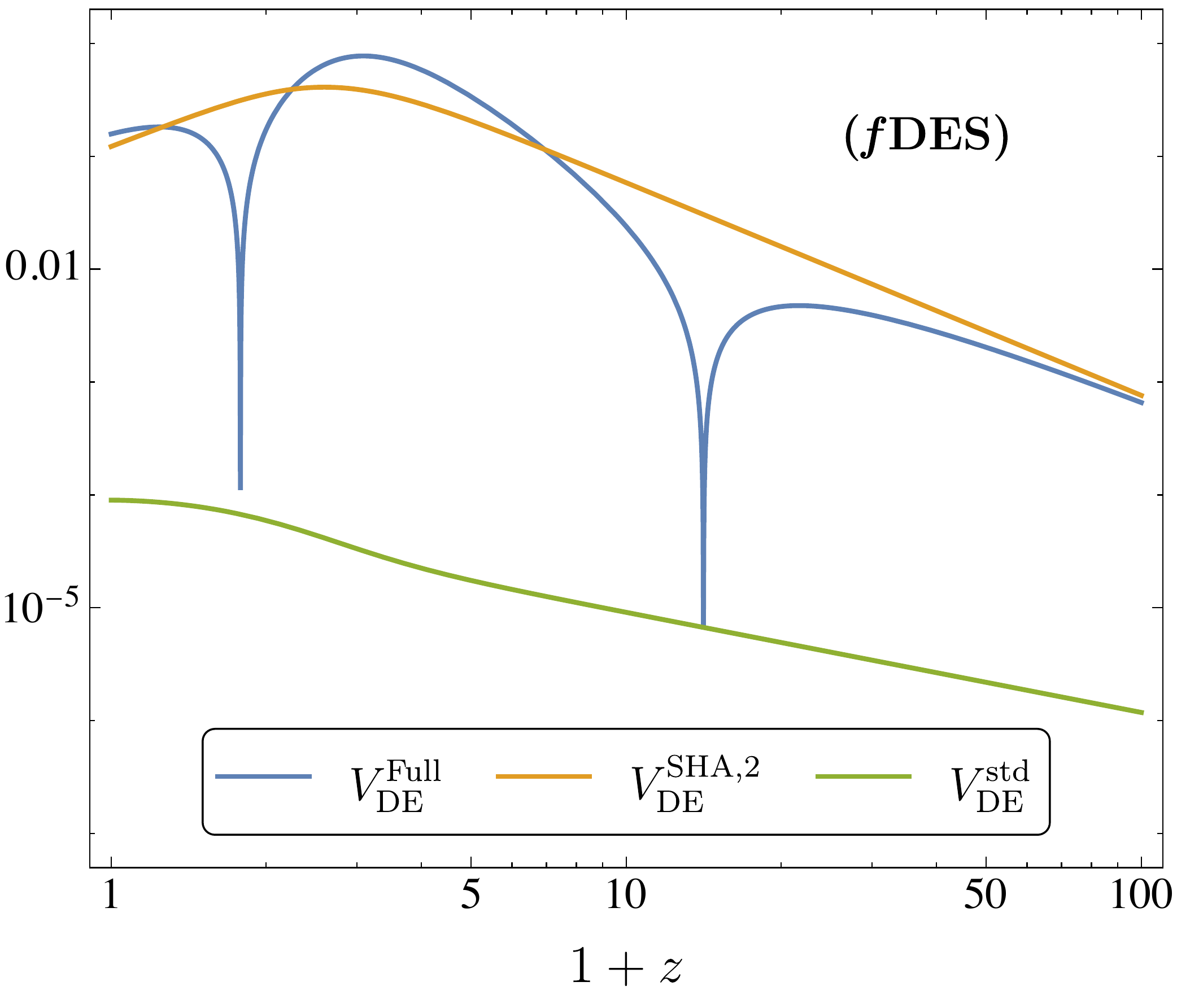}
\hfill
\includegraphics[width = 0.495\textwidth]{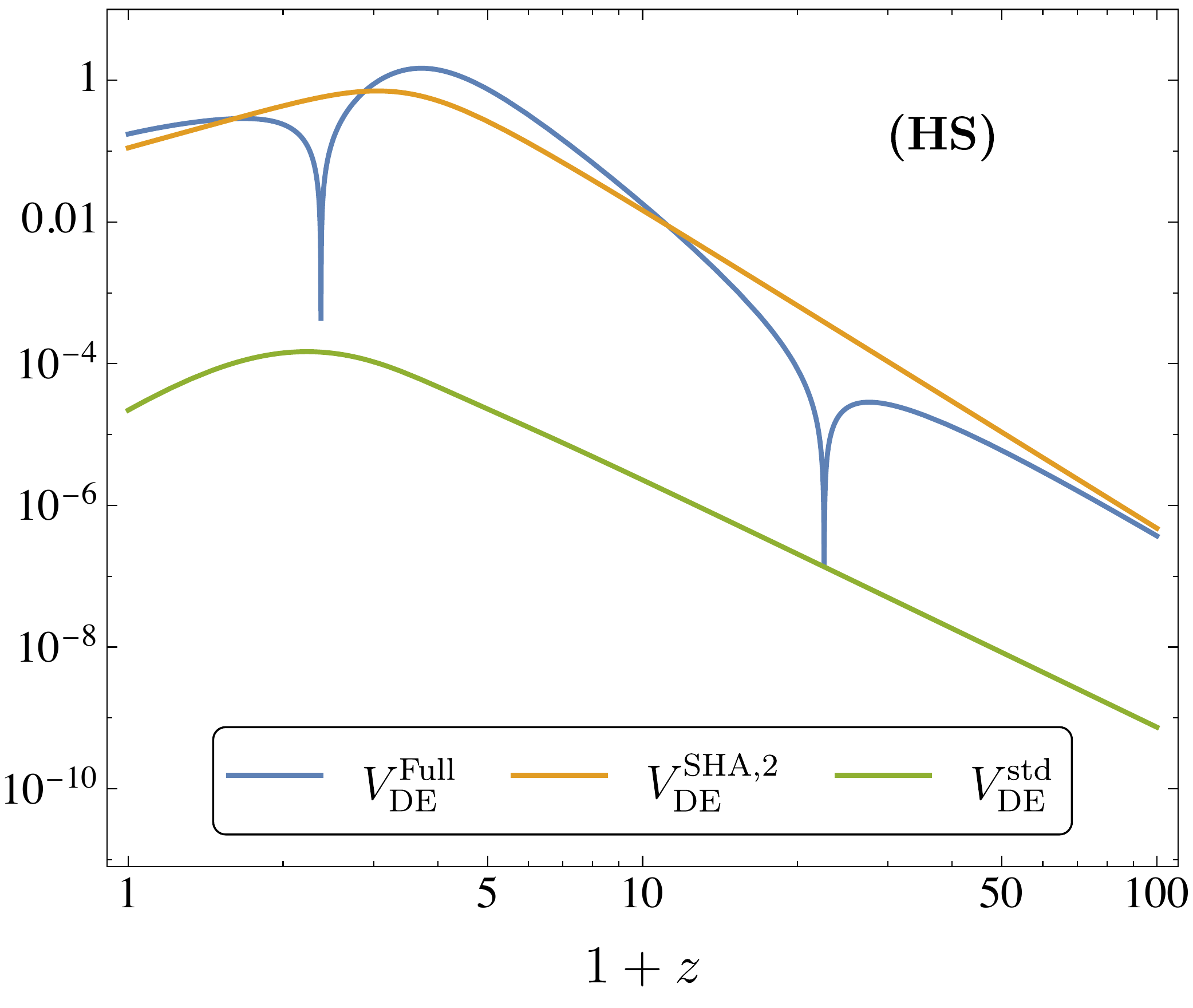}
\caption{Comparison of the DE scalar velocity obtained from full numerical solution, the standard QSA-SHA procedure, and our new parameterization. For both models ($f$DES on the left panel and HS on the right panel), we see that our expressions show overall agreement with the full $V_\text{DE}$, while the standard result is far smaller. This is due to $V^\text{std}_\text{DE}$ is a third-order expression in the new SHA parameterization.}
\label{Fig: VDE}
\end{figure}

From the results for the potentials, it is possible to compute the full DE perturbations using Eqs.~\eqref{Eq: deltaDE f(R)}-\eqref{Eq: stress f(R)}. As shown in Fig.~\ref{Fig: deltas}, the contribution from correction terms to the standard QSA-SHA results are hardly noticeable in the evolution of $\delta_\text{DE}$. Nonetheless, from the corresponding subplots, we can distinguish that some differences arise. In general, we define the percentage difference of an approximated quantity $X_\text{Approx}$ (whether it is obtained from the standard approach or our new approach) with respect to the corresponding full numerical solution $X_\text{Full}$ as
\begin{equation}
\text{Diff}\% = \left| \frac{X_\text{Full} - X_\text{Approx}}{X_\text{Full}} \right| \times 100.
\end{equation}
\begin{figure}
\includegraphics[width = 0.495\textwidth]{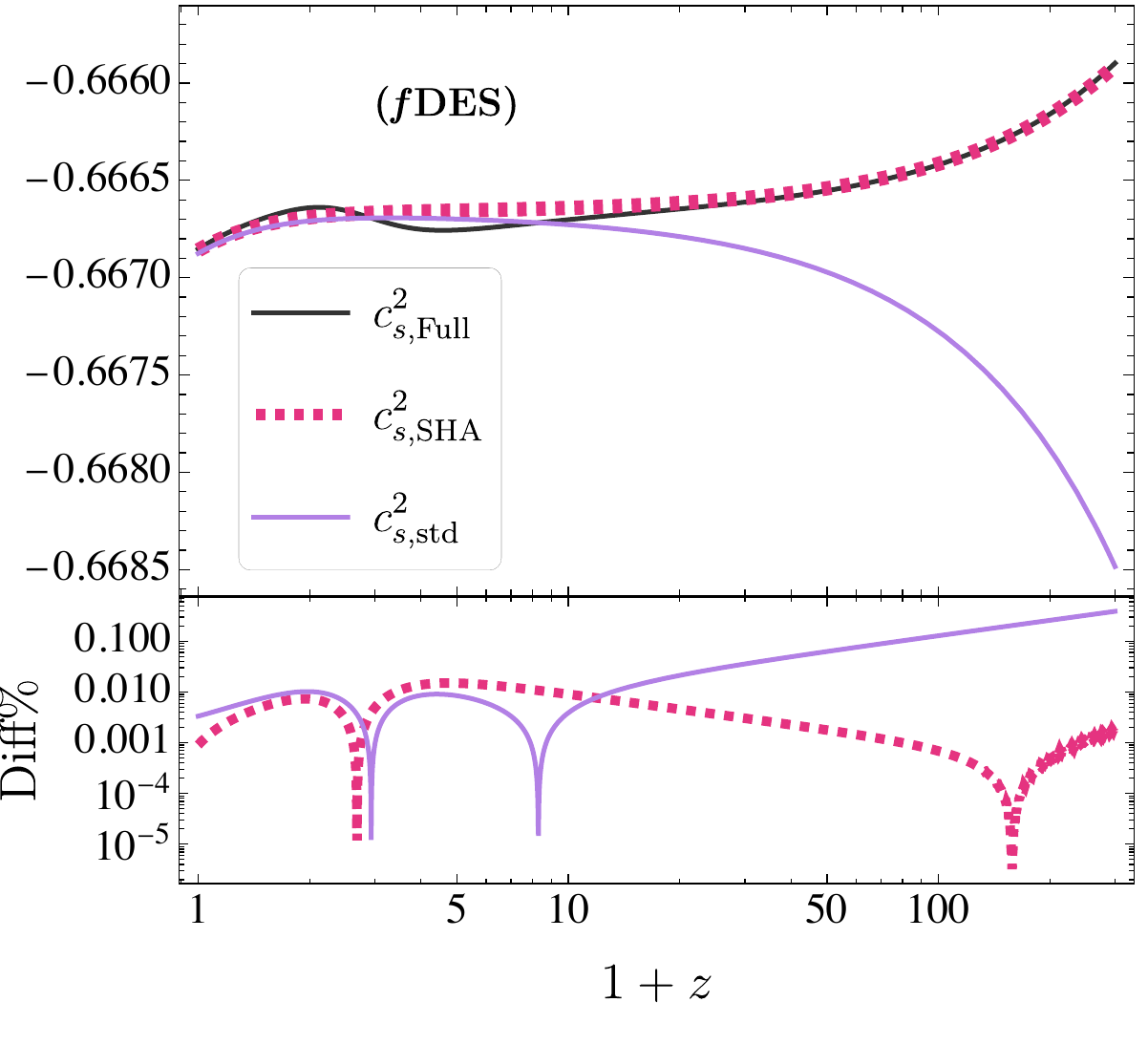}
\hfill
\includegraphics[width = 0.495\textwidth]{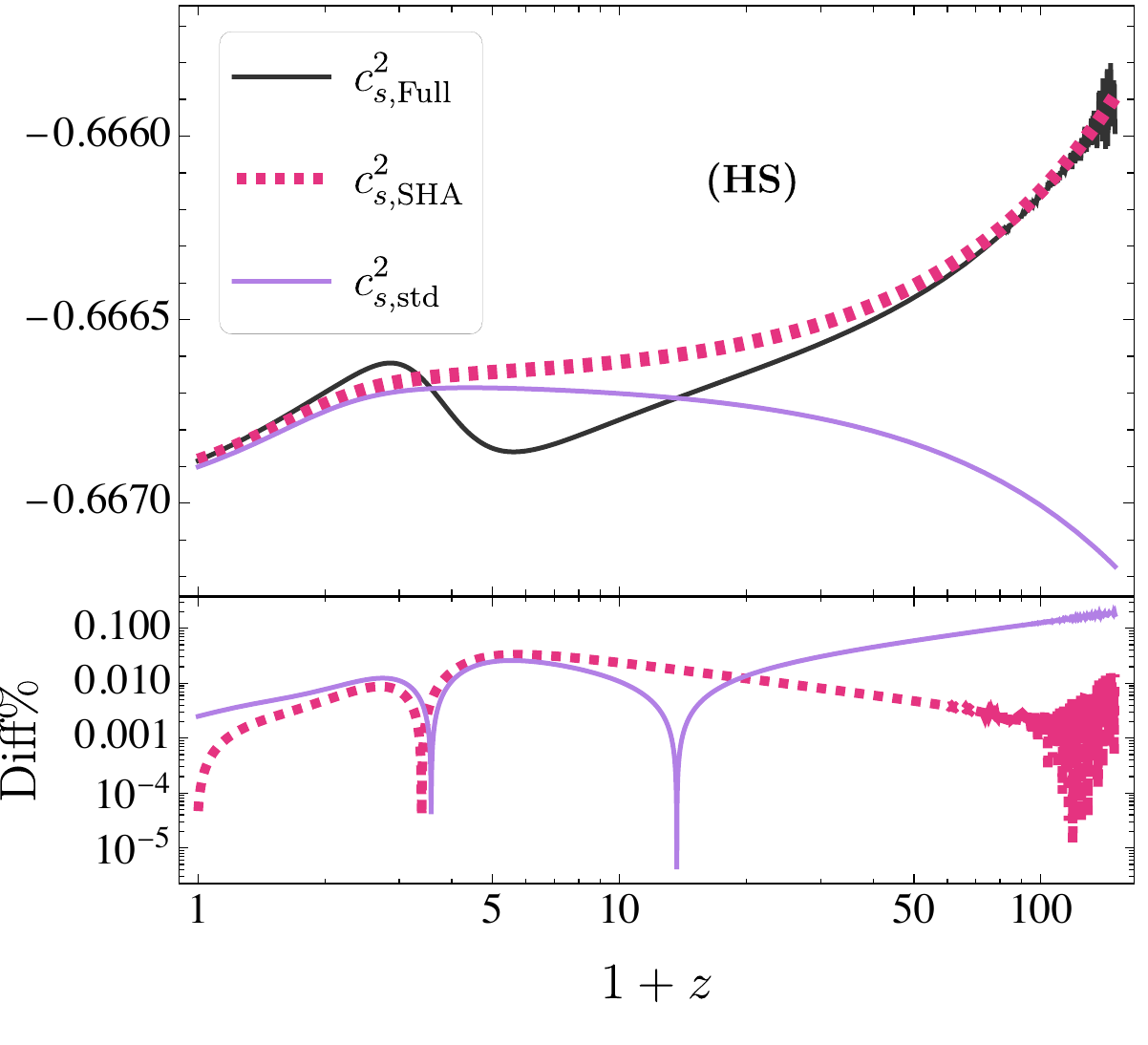}
\caption{Comparison of the DE sound speed obtained from full numerical solution, the standard QSA-SHA approach, and our new parameterization. For both models ($f$DES on the left panel and HS on the right panel), we see that both approaches give accurate results, however, our new scheme performs better at earlier times since the standard approach fails in accounting the decaying behavior of $c_{s, \text{DE}}^2$ from matter dominance. This difference is due to the existence of second-order terms neglected in $\delta P^\text{std}_\text{DE}$, which only considers fourth and sixth-order corrections in the new SHA parameterization.}
\label{Fig: Sound speed}
\end{figure}
It is noteworthy from the subplots in Fig.~\ref{Fig: deltas} that the simple extensions from our approach to the standard QSA-SHA results improve the accuracy in the description for $\delta_\text{DE}$ by around two orders of magnitude. On the other side, as expected from the discussion in Sec.~\ref{Sec: f(R) Theories}, there are significant differences between the DE scalar velocity given by the standard approach in Eq.~\eqref{Eq: std VDE} and our result at first-order in SHA parameters in Eq.~\eqref{Eq: New VDE}. In Fig.~\ref{Fig: VDE}, we note that $V_\text{DE}$ given by the standard QSA-SHA procedure is far smaller than the full $V_\text{DE}$, while our second-order expression shows overall agreement for both models. It is possible to track the origin of this disagreement. Note that our expression for $V_\text{DE}$ is a first-order expression in the $\ve$-expansion, however, if we use our parameterization in the standard $V_\text{DE}$ given in Eq.~\eqref{Eq: std VDE} we get
\begin{equation}
V^\text{std}_\text{DE} = \frac{3 a F_R}{F}\frac{F + 6 \frac{k^2}{a^2} F_R}{ F + 3 \frac{k^2}{a^2} F_R } (\delta + \xi - 2) \kappa \bar{\rho}_m \delta_m  \frac{k^3}{a^3} \ve^3,
\end{equation}
which clearly is a third-order expression in the new SHA parameterization. We verified that a similar problem arises in the case of the pressure perturbation, since the standard results for $\delta P_\text{DE}$ considers fourth and sixth-order corrections when $\ddot{F}$ is written in terms of $\ve$ [see Eqs.~\eqref{Eq: std dPDE} and \eqref{Eq: ddot F}], while our expression considers second-order corrections [see Eq.~\eqref{Eq: New deltaPDE}]. The impact of this difference can be seen in the evolution of the sound speed $c_{s, \text{DE}}^2 \equiv \delta P_\text{DE} / \delta \rho_\text{DE}$. In Fig.~\ref{Fig: Sound speed}, we can see that both methodologies provide accurate results for $c_{s, \text{DE}}^2$. However, note that the standard approach fails in accounting the general trend of the DE sound speed to decay from matter domination to present days. Instead, our new scheme shows perfect agreement with this behavior, and thus performs better at early times, as it can be seen in the subplots of Fig.~\ref{Fig: Sound speed}. To end this subsection, we would like to mention that in $f(R)$ theories, usually the sound speed is negative but the effective sound speed that also takes into account the DE anisotropic stress is always positive \cite{Cardona:2014iba}.

%%%%%%%%%%%%%%%%%%%%%%%%%%%%%%%
\subsubsection{Other Scales}
%%%%%%%%%%%%%%%%%%%%%%%%%%%%%%%

\begin{figure}[t!]
\includegraphics[width = 0.495\textwidth]{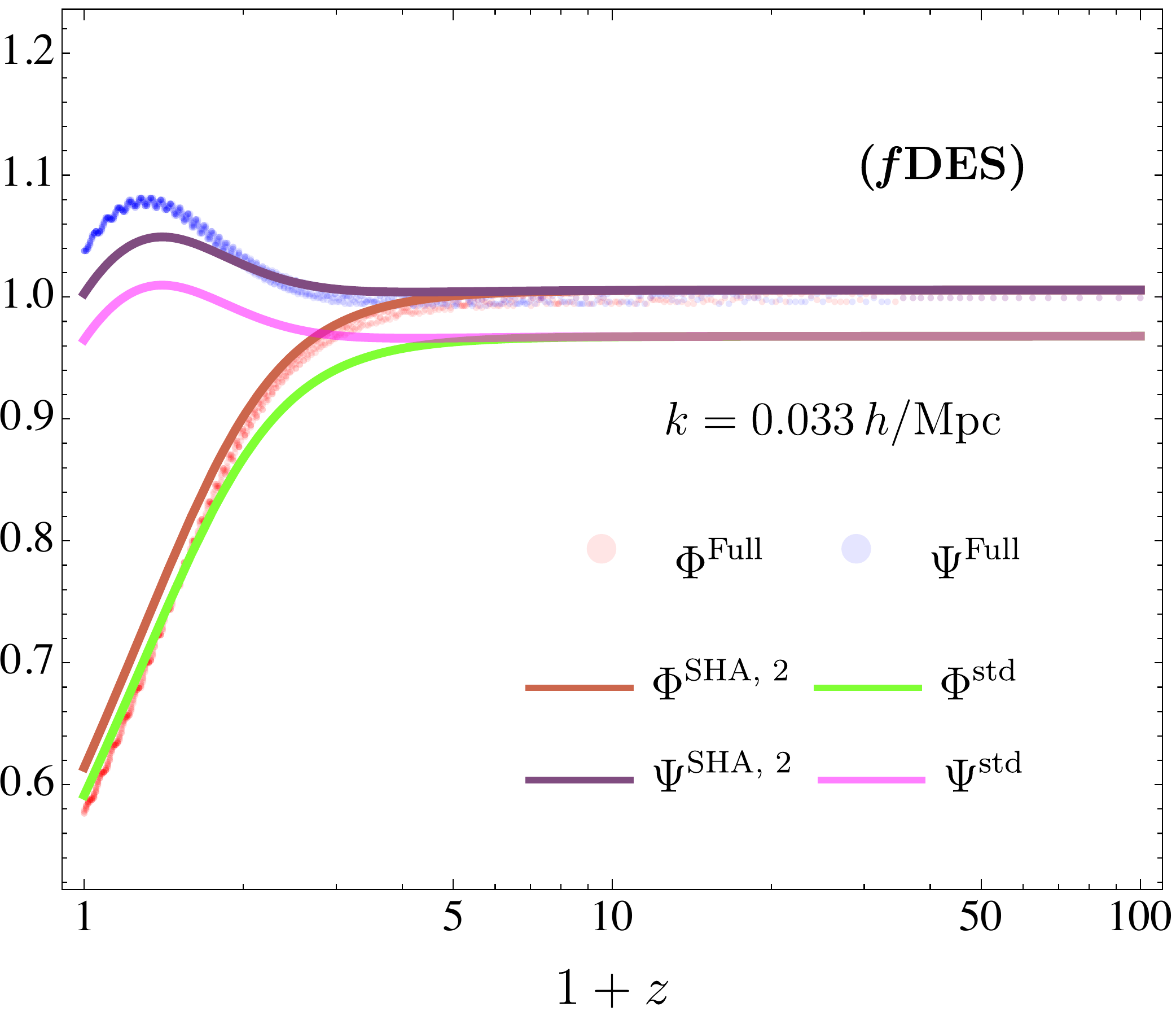}
\hfill
\includegraphics[width = 0.495\textwidth]{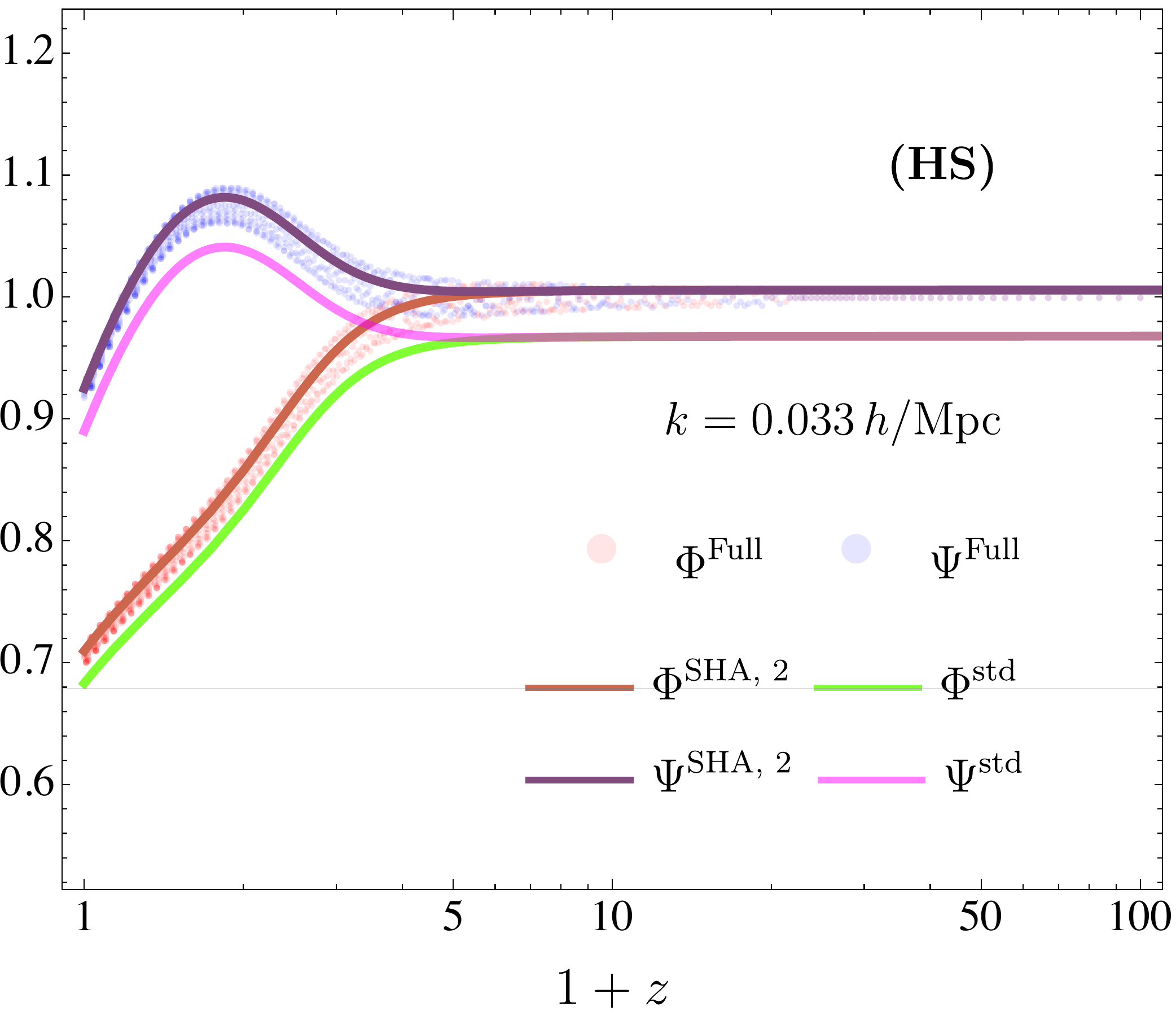}
\caption{Comparison of the potentials for the $f$DES model (left panel) and the HS model (right panel) considering a value for $k$ which is out of the ``safety region'', i.e., $k < 0.06 \, h/\text{Mpc}$. We see that the standard procedure provides less accurate results in contrast with the results from our approach, which is fairly acceptable for the HS model. However, in the case of the $f$DES model, results are not as accurate as in Fig.~\ref{Fig: Potentials}.}
\label{Fig: Potentials Scales}
\end{figure}

In order to test the ``safety region'' where the QSA and the SHA are applicable, namely $k \gtrsim 0.06 \, h/\text{Mpc}$, we compare the potentials obtained from the full solution considering $k = 0.033 \, h/\text{Mpc}$, and their corresponding results from the standard QSA-SHA procedure and our new approach. In the right panel Fig.~\ref{Fig: Potentials Scales}, we see that in the case of the HS model, our zero-order expansion in QSA parameters and second-order expansion in SHA parameters expressions still provides fairly accurate descriptions of the potentials, while the potentials obtained from the standard procedure have an offset diminishing the accuracy of these expressions. This offset in the standard QSA-SHA expressions is also present in the $f$DES model, as it can be seen in the left panel of Fig.~\ref{Fig: Potentials Scales}. Nonetheless, our new expressions also show deviations from the full solutions in this case. This indicates that both methods are not as accurate for $k = 0.033 \, h/\text{Mpc}$ (which is out of the safety region) as they are for $k = 0.1 \, h/\text{Mpc}$ (which is in the safety region), as can be seen in Fig.~\ref{Fig: Potentials}, and that the QSA parameters cannot be neglected anymore. As a final remark, we would like to mention that the ``safety region'' for the applicability of the QSA and the SHA, namely $0.06 \, h/\text{Mpc} \lesssim k \lesssim 0.2 \, h/\text{Mpc}$, is rather an estimation and not a general statement. The effectiveness of these approximations depends on the specific cosmological model and in this work we only studied models whose background are similar or equal to $\Lambda$CDM. However, we recall that $f(R)$ models with a very different background evolution in comparison to $\Lambda$CDM are tightly constrained by several tests \cite{Amendola:2006we, Pogosian:2007sw, Battye:2017ysh}. 

%%%%%%%%%%%%%%%%%%%%%%%%%%%%%%%%%%%%%%%%%%%%%%%%%%%
\section{Impact on Cosmological Observables}
\label{Sec: Impact on Cosmological Observables}
%%%%%%%%%%%%%%%%%%%%%%%%%%%%%%%%%%%%%%%%%%%%%%%%%%%

As we saw in the last section, some relevant terms are neglected in the DE perturbations when the standard QSA-SHA procedure is applied. So that, drastic modifications are taken into account for some quantities in our approach, e.g., the scalar velocity $V_\text{DE}$ is raised by roughly three orders of magnitude (see Fig.~\ref{Fig: VDE}), while some other quantities get minor corrections, such as the sound speed (see Fig.~\ref{Fig: Sound speed}). Therefore, it is of great interest to assess the influence of these changes on cosmological observables or parameters used in forecasts. In particular, we will analyze the impact of DE perturbations on: $i)$ the viscosity parameter which can be important in weak lensing experiments \cite{Sapone:2013wda}, and $ii)$ the matter power spectrum and the CMB angular power spectrum using the \texttt{EFCLASS} branch \cite{Arjona:2018jhh} of the Boltzmann solver \texttt{CLASS} \cite{Blas:2011rf}.

%%%%%%%%%%%%%%%%%%%%%%%%%%%%%%%%%%%%%%%%
\subsection{The Viscosity parameter}
%%%%%%%%%%%%%%%%%%%%%%%%%%%%%%%%%%%%%%%%

\begin{figure}[t!]
\includegraphics[width = 0.495\textwidth]{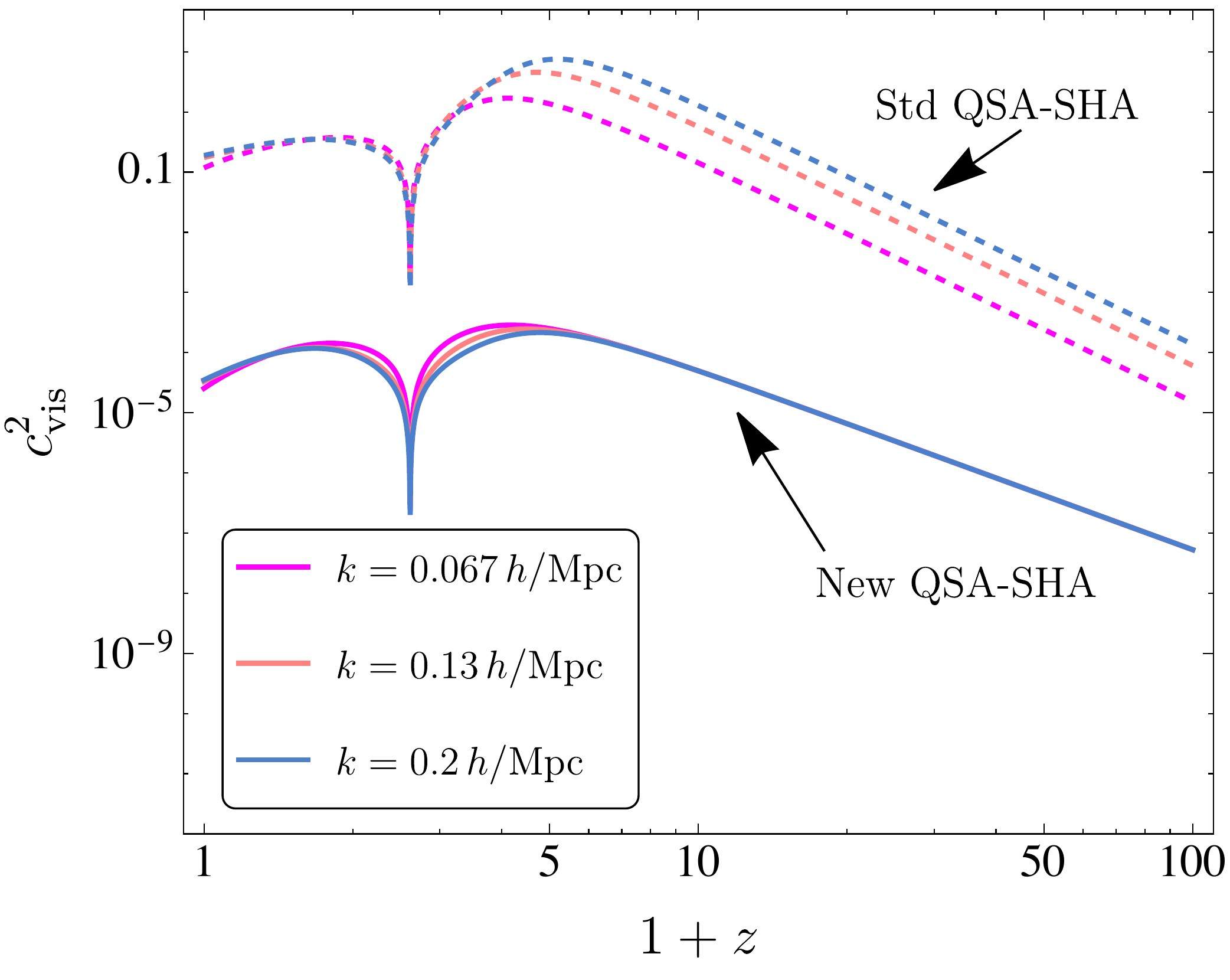}
\hfill
\includegraphics[width = 0.465\textwidth]{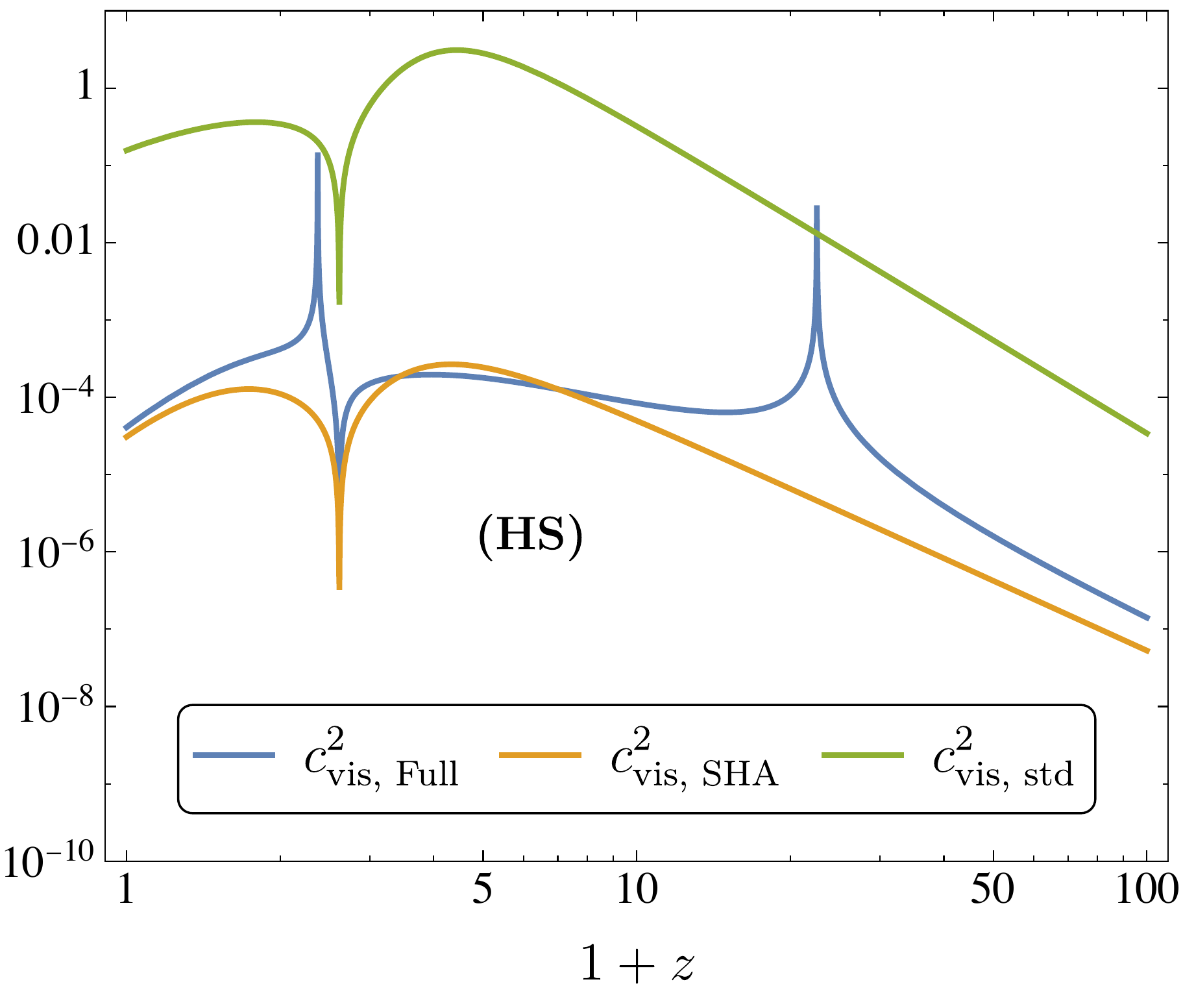}
\caption{$c^2_\text{vis}$ for the HS model. Left: Comparison between $c_\text{vis}^2$ computed from the standard approach and our new approach for three values of $k$. We see that the standard approach predicts a large viscosity parameter, while our method predicts a small one. Right: Comparison of $c_\text{vis}^2$ obtained from the standard approach, our new approach, and the full solution considering $k = 0.1 \, h/\text{Mpc}$. We see that the evolution obtained using our new SHA parameterization (mustard line) agrees in overall with the full numerical solution (blue line), while the result obtained from the standard QSA-SHA procedure (green line) gives a large over-estimation of this parameter.}
\label{Fig: Viscosity Parameter}
\end{figure}

In the $\Lambda$CDM model, it is well known that the potentials are related by $\Phi = - \Psi$. As it can be seen in Eq.~\eqref{Eq: Usual traceless Eq}, this is not the case in $f(R)$, since this difference can be related to the anisotropic stress as shown in Eq.~\eqref{Eq: stress f(R)}. Therefore, a detection of a non-zero anisotropic stress could favored MG theories over single-scalar field models which in general predicts a vanished $\pi_\text{DE}$ \cite{Kunz:2012aw, Saltas:2011xlz, Cardona:2014iba}. The effect of the anisotropic stress can be modelled by introducing a viscosity parameter that appears in the right hand-side of an effective DE anisotropic stress Boltzmann equation of the form \cite{Hu:1998kj}:
\begin{equation}
\dot{\sigma} + 3 H\frac{c_a^2}{w} \sigma = \frac{8}{3} \frac{c_\text{vis}^2}{(1 + w)^2} \frac{V_\mathrm{DE}}{a}, \label{eq:visc}
\end{equation}
where $c_a^2 = w - \frac{\dot{w}}{3 H (1 + w)} = w - \frac{a w'}{3(1 + w)}$ is the adiabatic sound speed. This can be inverted to solve for the viscosity parameter
\begin{equation}
c^2_\text{vis} = \frac{a H (1 + w_\text{DE})}{4w V_\text{DE}} \[ 3 c_{a, \text{DE}}^2 (1 + w_\text{DE}) \pi_\text{DE} + w\( a \pi'_\text{DE} - 3 w \pi_\text{DE} \) \],
\end{equation}
which is a function of the scale factor \cite{Hu:1998kj, Sapone:2013wda, Arjona:2018jhh}. From the last equation, it is clear that a large modification to $V_\text{DE}$ could strongly impact the evolution of the viscosity parameter. In the left panel of Fig.~\ref{Fig: Viscosity Parameter}, we compute $c_\text{vis}^2$ in the HS model considering three values of $k$ in the safety region, namely $0.067 \, h/\text{Mpc}$, $0.13 \, h/\text{Mpc}$, and $0.2 \, h/\text{Mpc}$. We can see that the viscosity parameter computed using the standard procedure reaches values greater than $1$ for $z \sim 10$, in contrast, this parameter computed using our new approach remains small. In the right panel, we verify that, for $k = 0.1 \, h/\text{Mpc}$, the parameter $c_\text{vis}^2$ computed using our new parameterization is in overall agreement with respect to the full solution, while the computation using the standard approach over-estimates this parameter by several orders of magnitude. In Ref. \cite{Sapone:2013wda}, it was found that a small viscosity, around $c^2_\text{vis} \sim 10^{-4}$ for the $w$CDM model, could be detected by Euclid when weak lensing and galaxy clustering data are combined. Although in Ref. \cite{Sapone:2013wda}, the viscosity parameter was taken as a constant, which is not realistic in a general dark energy scenario as was pointed out in Ref. \cite{Arjona:2018jhh}, if we take this constant value as a representative value of the smallness of $c^2_\text{vis}$, we see that our new parameterization indeed agrees with it.

%%%%%%%%%%%%%%%%%%%%%%%%%%%%%%%%%%%%%%%%
\subsection{Linear Power Spectra}
%%%%%%%%%%%%%%%%%%%%%%%%%%%%%%%%%%%%%%%%

\begin{figure}[t!]
\includegraphics[width = 0.495\textwidth]{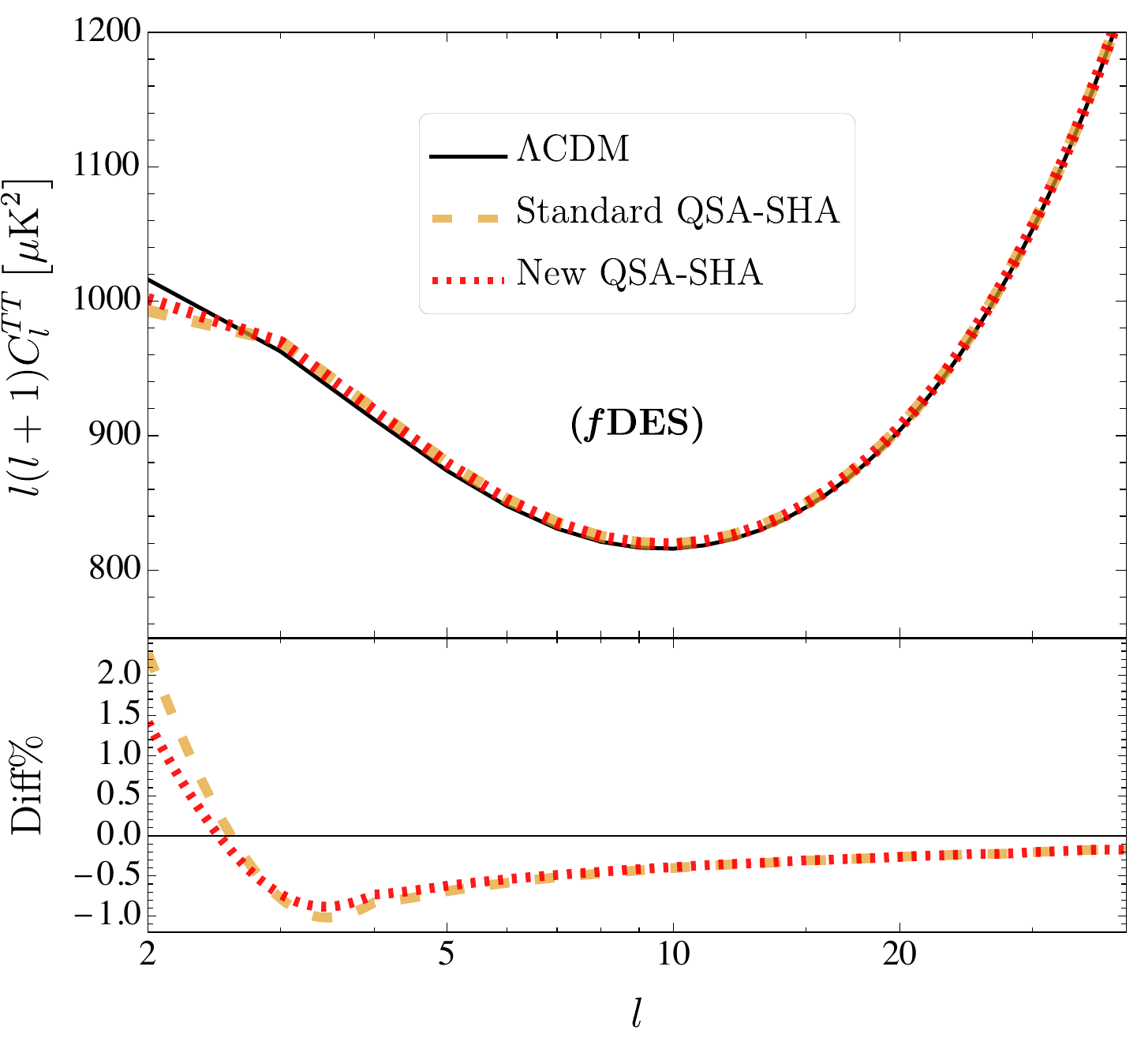}
\hfill
\includegraphics[width = 0.495\textwidth]{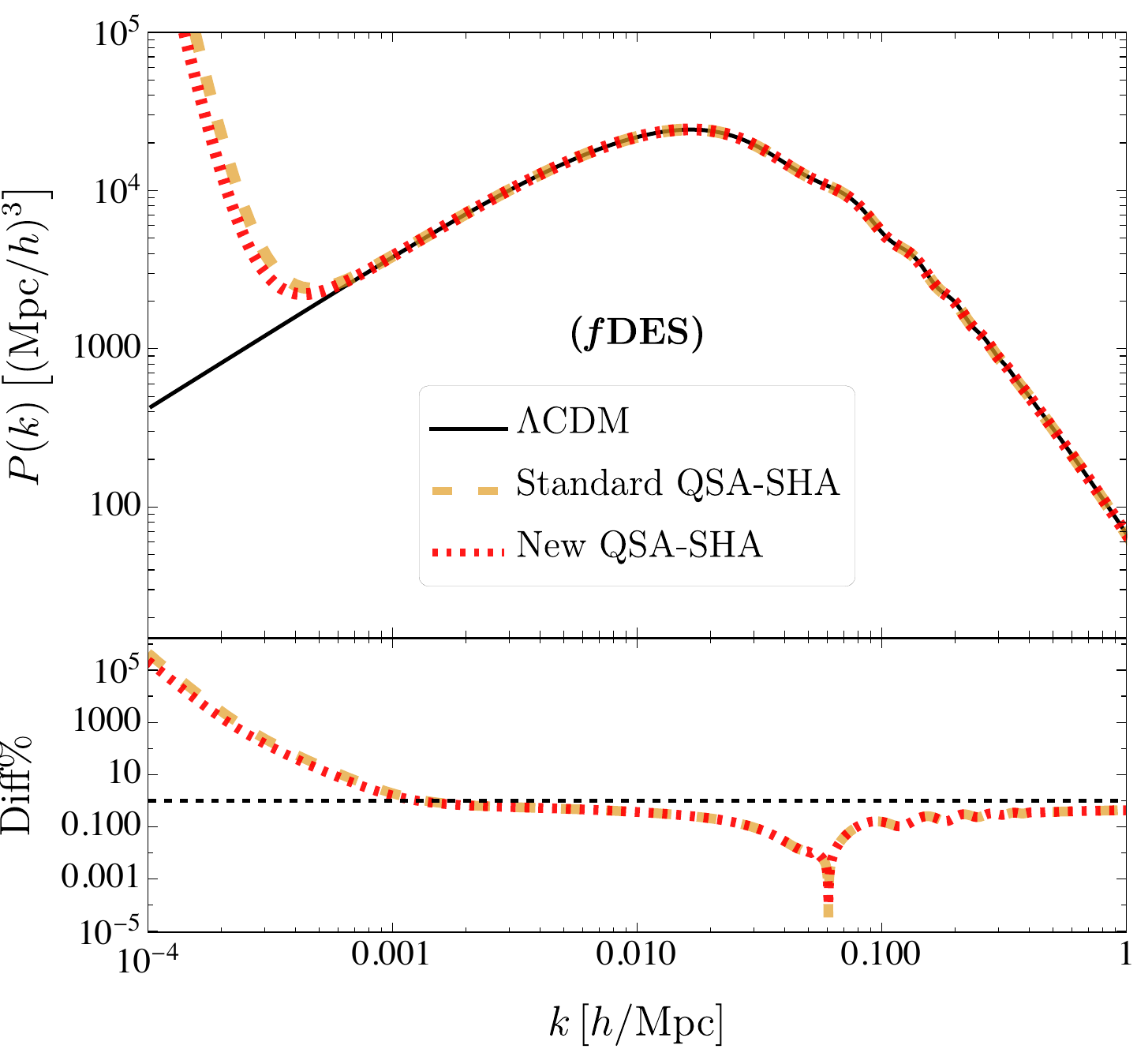}
\caption{$TT$ CMB power spectrum for low-$l$ multipoles (left) and matter power spectrum (right) in the $f$DES and $\Lambda$CDM models. We see that the results from both approximation schemes are practically indistinguishable for $l \gtrsim 5$ in the case of the CMB spectrum and for $k \gtrsim 0.001 \, h/\text{Mpc}$ in the case of $P(k)$. For lower multipoles, we see some minor differences, while for smaller values of $k$ both methods give divergent results, mainly due to that the SHA is not applicable.}
\label{Fig: Observables}
\end{figure}

In Fig.~\ref{Fig: Observables} we plot the low-$l$ multipoles of the CMB angular power spectrum (left), and the linear matter power spectrum (right) for the $f$DES model using the standard QSA-SHA procedure and our new approach. We plot the $\Lambda$CDM Planck best-fit 2018 standard result in both cases for comparison. In the case of the CMB power spectrum, we see that the $f$DES model do not apart significantly from the $\Lambda$CDM prediction, and only at the lowest multipoles ($l < 5$) both approximation scheme differ. However, this difference is not representative (less than $1\%$). For the matter power spectrum, we note that the standard approach and the new approach predictions for $P(k)$ are completely degenerate for modes $k \gtrsim 0.001 \, h/\text{Mpc}$. For smaller modes, or equivalently for larger scales, both schemes yield to divergent results. This divergent behavior is common in theories where the QSA-SHA are assumed (see e.g. Ref.~\cite{Cardona:2022lcz} where a similar divergency was found in the context of Scalar-Vector-Tensor theories \cite{Heisenberg:2018acv, Heisenberg:2018mxx}). As mentioned in previous sections, we do not expect the QSA and the SHA hold for modes smaller than $k < 0.067 \, h/\text{Mpc}$. Moreover, note that in the region $k \lesssim k_\text{eq}$, where $k_\text{eq}$ is the mode entering the horizon at the radiation-matter equilibrium, radiation is not negligible anymore. Therefore, the expansion history cannot be described using the Hubble parameter in Eq.~\eqref{Eq: Friedman Eqs} and the perturbations do not follow equations similar to Eqs.~\eqref{Eq: Einstein 00}-\eqref{Eq: Traceless Einstein ij}.

Notice that although our new QSA-SHA scheme provide more accurate descriptions of the DE perturbations than the standard approach, this improvement is not reflected in the linear spectra in Fig.~\ref{Fig: Observables}. We realized that $V_\text{DE}$ is a dimensionful quantity, and thus it has to be compared with quantities with the same dimensions. In the case of the linear observables, they are mainly affected by the comoving density perturbation defined as
\begin{equation}
\Delta_\text{DE} \equiv \delta_\text{DE} + \frac{3 a H}{k^2} V_\text{DE},
\end{equation}
which is a gauge invariant quantity. For sub-horizon modes, the term $H V_\text{DE}/k^2$ is completely negligible and $\Delta_\text{DE} \approx \delta_\text{DE}$. Since the standard approach indeed provides fairly accurate results for the potentials $\Phi$ and $\Psi$, and for the density contrast $\delta_\text{DE}$, it is no a surprise that the improvement on these quantities do not represent a major modification on the linear observables in the scales where these approximations are valid.

%%%%%%%%%%%%%%%%%%%%%%%%%%%%
\section{Conclusions}
\label{Sec: Conclusions}
%%%%%%%%%%%%%%%%%%%%%%%%%%%%
In this work, we have revisited the quasi-static and sub-horizon approximations when applied to $f(R)$ theories. We do so by proposing a parameterization using slow-rolling functions (like in inflation) which allowed us to track the relevance of time-derivatives of the gravitational potentials and the scale $k$ in contrast to the expansion rate $H$ [see Eqs.~\eqref{Eq: SHA parameters} and \eqref{Eq: QSA Parameters}]. In the effective fluid approach to $f(R)$, we showed that the standard QSA-SHA results found in the literature [Eqs.~\eqref{Eq: Standard Potentials}-\eqref{Eq: std Stress}] correspond to the 0 order expansion in our new scheme. In particular, we showed that second-order corrections can improve the description of the dark energy perturbations by several orders of magnitude in most cases, e.g., $\delta_\text{DE}$ (see Fig.~\ref{Fig: deltas}), and turned out to be significant for some DE effective fluid quantities. In  particular, we found large differences in the DE scalar velocity $V_\text{DE}$ (see Fig.~\ref{Fig: VDE}) and a minor correction in the sound speed $c_{s, \text{DE}}^2$ which increases the accuracy of the approximations during the matter dominated epoch (see Fig.~\ref{Fig: Sound speed}). The accuracy of our expressions [Eqs.~\eqref{Eq: No QSA Phi}-\eqref{Eq: New stress}], and the standard QSA-SHA results, were assessed by comparing with results obtained from a full numerical solution of the linear perturbation equations. We found general agreement between the results from our new expressions and the full numerical solution where no approximations are assumed.

We estimated a ``safety region'' where the QSA and the SHA can be applied. We found that the QSA and the second-order expansion in the SHA yield very accurate results from matter domination ($z \lesssim 100$) to the present ($z = 0$) whenever $k \gtrsim 0.067 \, h/\text{Mpc}$. We also showed for a particular value out of this region the loss of accuracy of these approximations (see Fig.~\ref{Fig: Potentials Scales}). This can also be seen in the matter power spectrum depicted in the right panel of Fig.~\ref{Fig: Observables}, where we can see that the curves diverge for $k$ small. We would like to stress that a similar divergent behavior was found in Ref. \cite{Cardona:2022lcz}. However, note that previous works on Horndeski theories \cite{Arjona:2019rfn}, and Scalar-Vector-Tensor theories \cite{Cardona:2022lcz}, where these approximations were used, all of them agree, in their corresponding limits, with the standard expressions in $f(R)$. In the case of the lowest multipoles of the CMB power spectrum, and in general for the matter power spectrum, the improvement in the accuracy from our expressions do not imply a corresponding improvement in these linear observables. This was expected since the QSA-SHA indeed provide accurate descriptions of the potentials $\Phi$ and $\Psi$, and the density contrast $\delta_\text{DE}$. In the particular case of the scalar velocity $V_\text{DE}$, the large difference between the standard prediction and our new prediction is not relevant for these observables, since the term $V_\text{DE}/k^2$ is completely sub-dominant for sub-horizon scales. 

We want to stress that the main goal of this work was to provide a transparent framework to track the relevance of each term in the perturbation equations, in order to clarify the application of the QSA-SHA when applied to modified gravity theories. To conclude, clearly our results warrant performing a more in-depth analysis of the validity of the QSA and the SHA approximations in more generalized MG theories, e.g., of the Horndeski type, however we leave this for future work.

\section*{Acknowledgements}
Several codes with the main results presented here can be found in the GitHub repository \href{https://github.com/BayronO/QSA-SHA-for-Modified-Gravity}{https://github.com/BayronO/QSA-SHA-for-Modified-Gravity} of BOQ. The authors are grateful to Rub\'en Arjona, Wilmar Cardona, and C\'esar Valenzuela-Toledo for useful discussions. BOQ would like to express his gratitude to the Instituto de Física Téorica UAM-CSIC, in Universidad Autonóma de Madrid, for the hospitality and kind support during early stages of this work. BOQ is also supported by Patrimonio Aut\'onomo - Fondo Nacional de Financiamiento para la Ciencia, la Tecnolog\'ia y la Innovaci\'on Francisco Jos\'e de Caldas (MINCIENCIAS - COLOMBIA) Grant No. 110685269447 RC-80740-465- 2020, projects 69723 and 69553. SN acknowledges support from the research project  PID2021-123012NB-C43, and the Spanish Research Agency (Agencia Estatal de Investigaci\'on) via the Grant IFT Centro de Excelencia Severo Ochoa No CEX2020-001007-S, funded by MCIN/AEI/10.13039/501100011033.

\appendix

%%%%%%%%%%%%%%%%%%%%%%%%%%%%%%%%%%%%%%%%%%%%%%%
\section{Linear Field Equations in $f(R)$}
\label{App: Linear Field Equations in $f(R)$}
%%%%%%%%%%%%%%%%%%%%%%%%%%%%%%%%%%%%%%%%%%%%%%%
Here we present the expressions for the coefficients in Eqs.~\eqref{Eq: 00 f(R)}-\eqref{Eq: Traceless f(R)}:\begin{equation}
A_1 = 3\dot{F} + 3 F H, \qquad A_2 = - 3 F H, \qquad A_3 = 3 F \dot{H} + 3F H^2 - 3 H \dot{F},
\end{equation}
\begin{equation*}
A_4 = F, \qquad A_5 = 3 F \dot{H} - 3F H^2 - 9 H \dot{F}, \qquad A_6 = -F.
\end{equation*}

\begin{equation}
C_1 = - F, \qquad C_2 = F, \qquad C_3 = \dot{F} - F H, \qquad C_4 = 2 \dot{F} + F H.
\end{equation}

\begin{equation}
B_1 = 3 F, \qquad B_2 = -3 F, \qquad B_3 = 12 F H, \qquad B_4 = -12 F H - 9 \dot{F},
\end{equation}
\begin{equation*}
B_5 = 3 F \dot{H} + 9 F H^2 - 6 H \dot{F} - 3 \ddot{F}, \qquad B_6 = - 9 F \dot{H} -9 F H^2 - 18 H \dot{F} - 9 \ddot{F}.
\end{equation*}

\begin{equation}
D_1 = 3, \qquad D_2 = 12 H, \qquad D_3 = - 3 H, \qquad D_4 = \frac{F}{2 F_R},
\end{equation}
\begin{equation*}
D_5 = 2, \quad D_6 = \frac{F}{2 F_R} - 6 \dot{H} - 12 H^2, \qquad D_7 = 1.
\end{equation*}

%%%%%%%%%%%%%%%%%%%%%%%%%%%%%%%%%%%%%%%%%%%%%%%%%%%
\section{Dark Energy Perturbations in $f(R)$}
\label{App: Dark Energy Perturbations in $f(R)$}
%%%%%%%%%%%%%%%%%%%%%%%%%%%%%%%%%%%%%%%%%%%%%%%%%%%
Here we present the expressions for the coefficients in Eqs.~\eqref{Eq: deltaDE f(R)}-\eqref{Eq: VDE f(R)}:
\begin{equation}
W_1 = 6 H - 3F H - 3 \dot{F}, \qquad W_2 = 3 F H, \qquad W_3 = -3 F \dot{H} - 3 F H^2 + 3 H \dot{F},
\end{equation}
\begin{equation*}
W_4 = 2 - F, \qquad W_5 = 3 H^2F - 6 H^2 2 - 3 F \dot{H} + 9 H \dot{F}, \qquad W_6 = F,
\end{equation*}

\begin{equation}
Y_1 = F - 2, \qquad Y_2 = -F, \qquad Y_3 = 4 F H - 6 H, \qquad Y_4 = 2H - 4F H - 3\dot{F},
\end{equation}
\begin{equation*}
Y_5 = 3 F H^2 + F \dot{H} - 2 H \dot{F} - \ddot{F}, \qquad Y_6 = - \frac{2}{3}, 
\end{equation*}
\begin{equation*}
Y_7 = 6 H^2 - 3 F H^2 + 4 \dot{H} - 3 F \dot{H} - 6 H \dot{F} - 3 \ddot{F}, \qquad Y_8 = - \frac{2}{3},
\end{equation*}

\begin{equation}
Z_1 = F - 2, \qquad Z_2 = - F, \qquad Z_3 = F H - \dot{F}, \qquad Z_4 = 2 H - F H - 2 \dot{F}.
\end{equation}

%%%%%%%%%%%%%%%%%%%%%%%%%%%%%%%%%%%%%%%%%%%%%%%%%%%%%%%%%%%%%%%%
\section{Linear Field Equations in the New Parameterization}
\label{App: Linear Field Equations in the New Parametrization}
%%%%%%%%%%%%%%%%%%%%%%%%%%%%%%%%%%%%%%%%%%%%%%%%%%%%%%%%%%%%%%%%
Here we present the expressions for the coefficients in Eqs.~\eqref{Eq: New 00 eq}-\eqref{Eq: New traceless eq}:
\begin{equation}
\mc{A}_1 = - \mc{A}_4 = F, \qquad \mc{A}_2 = 3 F (\delta + \ve_\Phi), \qquad \mc{A}_3 = 18 F_R (\ve_\Phi - 1) (\delta + \xi - 2),
\end{equation}
\begin{equation*}
\mc{A}_5 = 3 F (\delta - \ve_\Psi - 2), \qquad \mc{A}_6 = -54 F_R (\delta + \xi - 2),
\end{equation*}

\begin{equation}
\mc{D}_1 = \mc{D}_4 = \frac{F}{2 F_R}, \qquad \mc{D}_2 = 2, \qquad \mc{D}_3 = 3 \ve_\Phi (3 + \delta + \ve_\Phi + \chi_\Phi),
\end{equation}
\begin{equation*}
\mc{D}_5 = 1, \qquad \mc{D}_6 = -3 (2 + 2\delta + \ve_\Psi).
\end{equation*}

%%%%%%%%%%%%%%%%%%%%%%%%%%%%%%%%%%%%%%%%%%%%%%%%%%
%\section{Full Expression for the Potentials}
%\label{App: Full Expression for the Potentials}
%%%%%%%%%%%%%%%%%%%%%%%%%%%%%%%%%%%%%%%%%%%%%%%%%%

\bibliographystyle{JHEPmodplain}
\bibliography{paper}

\end{document}